\documentclass[%
 aip,
 rsi,
 amsmath,amssymb,
 reprint,%
]{revtex4-1}

\usepackage{graphicx}
\usepackage{dcolumn}
\usepackage{bm}

\usepackage[utf8]{inputenc}
\usepackage[T1]{fontenc}
\usepackage{mathptmx}
\usepackage{etoolbox}

\usepackage{multirow}
\usepackage{graphicx, color}
\usepackage{url}

\usepackage{hyperref}

\newcommand{\sslash}{\mathbin{/\mkern-6mu/}}
\newcommand{\TwoPi}{\ensuremath{360^{\circ}}}

\newcommand{\chg}[1]{#1}
\newcommand{\del}[1]{}

\makeatletter
\def\@email#1#2{%
 \endgroup
 \patchcmd{\titleblock@produce}
  {\frontmatter@RRAPformat}
  {\frontmatter@RRAPformat{\produce@RRAP{*#1\href{mailto:#2}{#2}}}\frontmatter@RRAPformat}
  {}{}
}%
\makeatother

\begin{document}

\preprint{AIP/123-QED}

\title{A flexible control system for atomic, molecular and optical physics experiments}

\author{A. Trenkwalder} 
\email{trenkwalder@lens.unifi.it}
\affiliation{Istituto Nazionale di Ottica del Consiglio Nazionale delle Ricerche (INO-CNR), 50019 Sesto Fiorentino, Italy}
\affiliation{European Laboratory for Nonlinear Spectroscopy (LENS), 50019 Sesto Fiorentino, Italy}

\author{M. Zaccanti} 
\affiliation{Istituto Nazionale di Ottica del Consiglio Nazionale delle Ricerche (INO-CNR), 50019 Sesto Fiorentino, Italy}
\affiliation{European Laboratory for Nonlinear Spectroscopy (LENS), 50019 Sesto Fiorentino, Italy}

\author{N. Poli} 
\email{poli@lens.unifi.it}
\affiliation{Istituto Nazionale di Ottica del Consiglio Nazionale delle Ricerche (INO-CNR), 50019 Sesto Fiorentino, Italy}
\affiliation{European Laboratory for Nonlinear Spectroscopy (LENS), 50019 Sesto Fiorentino, Italy}
\affiliation{Dipartimento di Fisica e Astronomia and INFN Sezione di Firenze,  Università degli Studi di Firenze, Via Sansone 1, 50019 Sesto Fiorentino, Italy}
%


\date{\today}

\begin{abstract}
We have implemented a control system for experiments in atomic, molecular and optical physics based on a commercial low-cost board,
featuring a field-programmable gate array as part of a system-on-a-chip on which a Linux operating system is running. The board features Gigabit Ethernet, allowing for fast data
transmission and operation of remote experimental systems. A single board can control a set of devices generating digital, analog and radio frequency signals with a precise timing given either by an external or internal clock. Contiguous output and input sampling rates of up to 40 MHz are achievable. Several boards can run synchronously with a timing error approaching 1\,ns.
For this purpose, a novel auto-synchronization scheme is demonstrated, with possible application in complex distributed experimental setups
with demanding timing requests.
\end{abstract}

\maketitle

\section{Introduction}

Experimental control and data acquisition systems are widespread in many fields of scientific and industrial research where test and measurement systems need to be controlled and experimental data have to be gathered.
For the application of controlling experiments in the field of atomic, molecular and optical (AMO) physics digital pulses, analog, radio and microwave frequency signals need to be generated at well-defined times.
For instance, laser cooling and trapping of atomic gases down to ultralow temperatures typically require a temporal resolution of one microsecond. For this task, field programmable gate arrays (FPGAs) are very well suited. These can generate arbitrary digital pulses which can be used to program digital-to-analog converters (DAC), direct-digital synthesizers (DDS), and other devices with the requested timing resolution. As a result, FPGAs are already successfully employed in both commercial \cite{NI-FPGA} and open source \cite{Artiq2007} control systems. 

Owing to their flexibility, FPGAs also find application for a wide range of different tasks, encompassing clock signal generation \cite{Ketterle2013}, DDS programming \cite{Meschede2015,Katori2015,Lu2017}, arbitrary waveform generation \cite{Hobson2019}, lock-in demodulation \cite{LIF2018}, high-speed data acquisition (DAQ) \cite{Tokamak2018}, digital feedback servo system \cite{Sias2018,Madison2018}. Moreover, FPGAs are increasingly used for the control of quantum systems and processors and as feedback devices for quantum measurements, and can be even used within cryogenic environments \cite{QuantumFeedback2013, cryogenic2016, cryogenic2017, SpinControl2020, QubiC2021}. Applications of FPGAs in space are becoming of growing interest \cite{FPGA_in_space}.
Despite of all of these applications, the development of a custom FPGA-based system is time consuming and commercial solutions tend to be expensive. Nonetheless, the advent of cheap, multi-purpose FPGA development boards targeted for hobbyists and students, offers a solution with low-cost and short development time, from which also experimental research can benefit thanks to the impressive capabilities of these boards.

Here we present a control system with a novel approach based on a commercial, low-cost system-on-a-chip (SoC) board, consisting of a central processing unit (CPU) which is tightly connected to an FPGA and to a set of hardware interfaces used to communicate with external devices. 
A Linux operating system, executed on the CPU, gives the flexibility to use high-level programming languages, which can be quickly adapted to any specific request, such as interfacing with external devices like USB, Secure Digital (SD) memory card or Ethernet with no need of additional hardware or specifically designed micro-controllers. 
Furthermore, the presence of an electrically isolated Gigabit Ethernet interface, allows fast data transfer and easy connection to remote locations. 

All these features represent a clear advantage of FPGA-SoC systems with respect to previously realized FPGA-based solutions \cite{Prevedelli2020}, not only in terms of the superior data rates offered by the Ethernet interface, but also by the additional flexibility given by the presence of the easy programmable CPU and the fact that these are stand-alone systems which can be utilised independently on the hardware and software environment.

As a powerful simple application of such extended capabilities, here we demonstrate a novel scheme to auto-synchronize several boards using only two coaxial cables and the Ethernet communication. Without user interaction or dedicated real-time networking hardware \cite{WhiteRabbit}, the propagation delays of the signals among distant boards are measured by the boards and are corrected automatically with a residual timing error approaching 1\,ns. 

The paper is organized as follows: First, we present the board architecture in Sec. \ref{sec:hardware}, and the developed software in Sec. \ref{sec:software}. We then present the measured performance and the auto-synchronization scheme in Sec. \ref{sec:results}, and discuss the results in Sec. \ref{sec:conclusions}.

\section{Hardware architecture} \label{sec:hardware}

\begin{figure}[t]
\begin{center}
\includegraphics[width=\columnwidth]{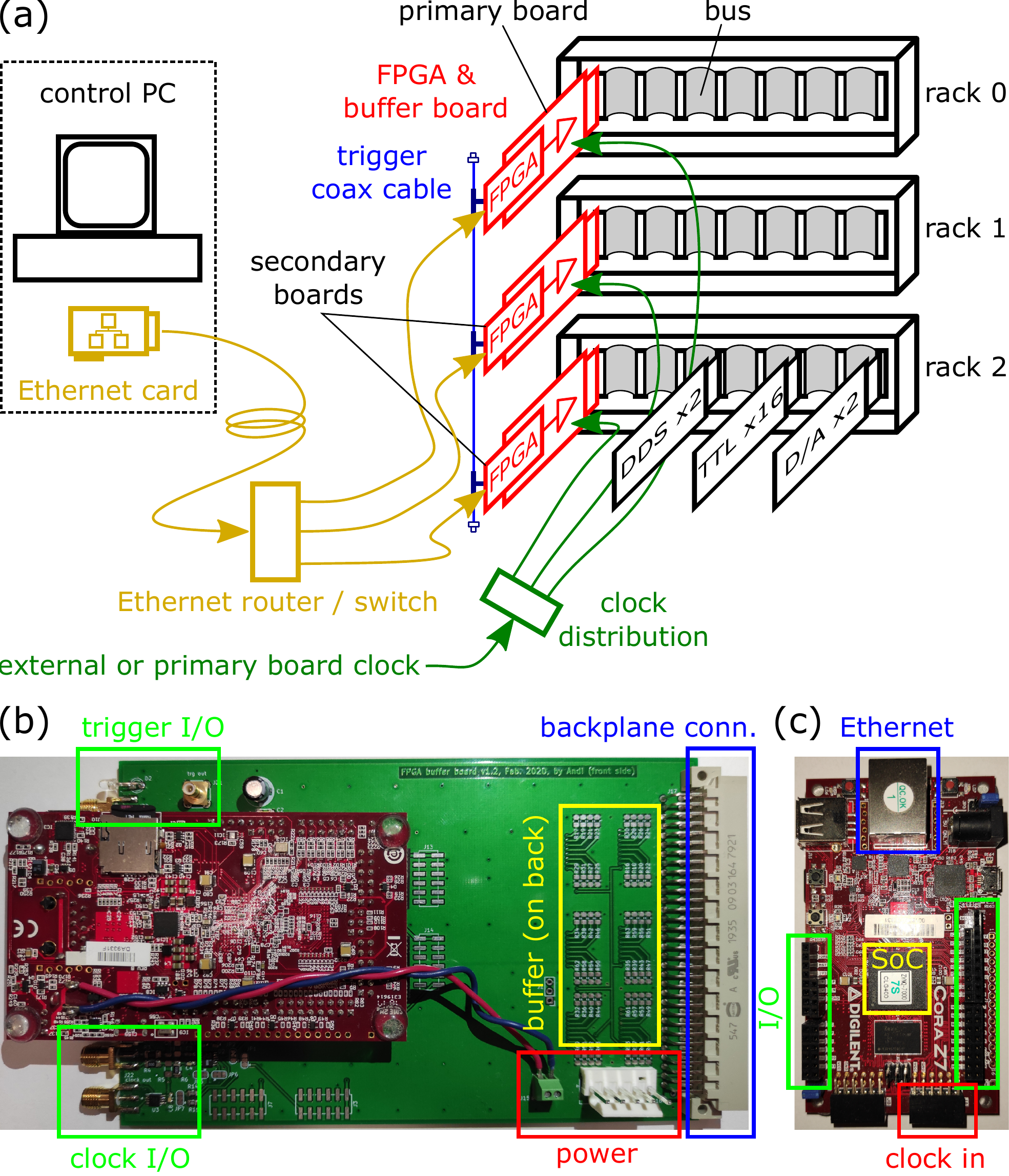}
\caption{a) Schematics of the control hardware. The experiment control sequence is sent from the control computer over an Ethernet network (yellow connections) to the FPGA-SoC boards (red). Each board, hosted in separated racks, where digital, analog and DDS devices can be freely inserted, is connect via buffer cards to a bus (gray ribbon cable). All FPGA-SoC boards are clocked (green connections) either by an external clock source or by the primary board clock signal. All the boards are synchronized via the clock and the trigger (blue connections) signals. b) Image of the FPGA-SoC board (red; back side visible), mounted on the buffer card (green; 100\,mm $\times$ 160\,mm Eurocard size). Backplane and power connectors are located on the right and bottom side. The trigger and clock I/Os are on the left-top and left-bottom side respectively. c) Image of the FPGA-SoC board (front side). The SoC is located in the center, the Ethernet connector is on the top side, and the two rows of pin sockets on the left and right side are used to connect with the buffer card. The external clock input is on one of the connectors on the bottom.}
\label{fig:setup}
\end{center}
\end{figure}

An overview of our setup is presented in Fig. \ref{fig:setup}a. A control computer generates the experiment control sequence (represented by a list of actions to be executed at a precise time) which is sent over Ethernet to one or several FPGA-SoC boards (distinguished by their IP address). Each FPGA-SoC board, hosted within a 19'' rack, drives via a buffer card a parallel bus over which digital and analog output devices and DDS are programmed at the specified time. These devices ultimately control the experiment and all physical parameters. 
The system is compatible with the well-established architecture in use at LENS, consisting of 
digital output devices with 16 TTL channels, analog output devices with two channels with 16-bit DACs with maximum $\pm$10\,V output, and DDS devices with two channels with up to 200\,MHz output frequency, which can be modulated in frequency and amplitude.
After the user has uploaded the control sequence, the experiment starts and the FPGA-SoC consecutively puts the data on the bus at the time defined in the time-stamp part of the control sequence. Once all samples are generated, the entire sequence can be repeated several times. For better timing accuracy, the clock source of the FPGA-SoC can be switched from the internal crystal oscillator to an externally provided clock signal. 

The heart of our control system is the Cora-Z7 board from Digilent \cite{Cora}, which hosts the Zynq-7007S (Zynq-7010) FPGA-SoC from Xilinx with a single (dual) core CPU (ARM Cortex A9) clocked at 650\,MHz. This represents the smallest FPGA-SoC from the Xilinx Zynq-7000 series. The board is provided with 512\,MB of DDR3 SDRAM (16\,bits data clocked at 525\,MHz) with Gigabit Ethernet and USB host and device ports. The FPGA part is nearly the same for the two variants and is similar to the low-end Artix-7 FPGA series, aiming for low-cost, low-power consumption and less demanding applications. It should be noted, that, while we choose a particular FPGA-SoC board with Gigabit Ethernet to implement our control system, the system and the methods presented in this paper can be implemented with any other FPGA-SoC boards with similar performance. For example, the DE10-Nano from Terasinc Inc. is a possible alternative\cite{TerasicDE10Nano}.

A custom designed buffer card \cite{github} is used to buffer the FPGA-SoC board signals and to shift the voltage level from the internal 3.3\,V logic level to the 5\,V (TTL) level of the bus. The buffer card also provides the needed buffers for the clock and trigger line used for the synchronization of different boards, as described below. An image of the FPGA-SoC board mounted on the buffer card is shown in Fig. \ref{fig:setup}b, and in Fig. \ref{fig:setup}c an image of the FPGA-SoC board (front side) is shown.

\subsection{The FPGA logic}\label{sub:logic}

\begin{figure}[t]
\begin{center}
\includegraphics[width=\columnwidth]{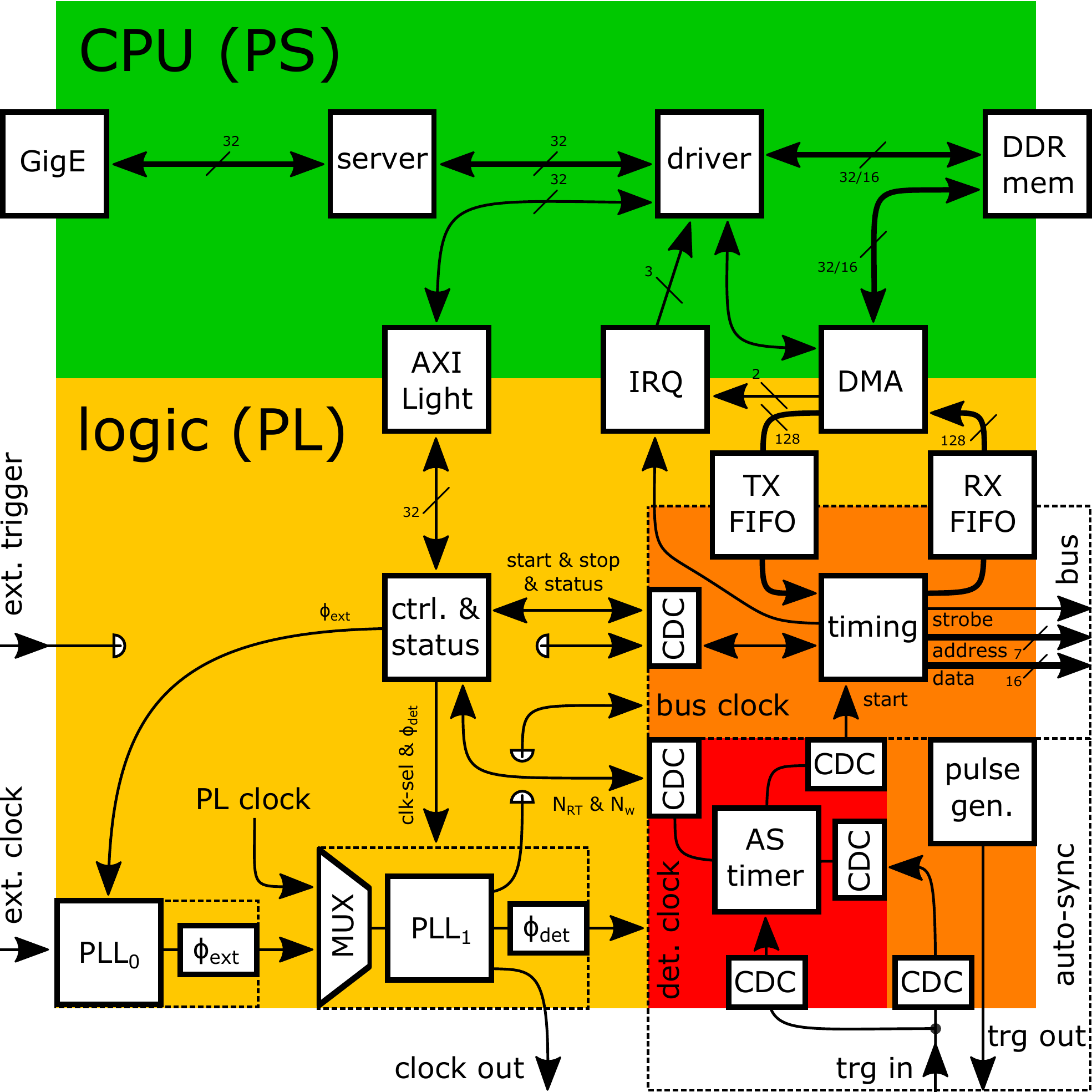}
\caption{Simplified block diagram of the Zynq-7000 SoC with the user data flow on the chip (thick lines). The processing system (PS, green) with 32-bit dual-core CPU allows the server and driver to access periphery like Gigabit Ethernet (GigE) and DDR memory using high-level programming languages and Linux system services. The programmable logic (PL, yellow/orange/red) holds the custom implementation of hardware. Interfaces efficiently transfer data between the two parts. Two phase-locked loop (PLL) modules are generating three different clocks (clock out, bus clock and detection clock) from an external clock source or from the PL system clock (yellow; selected by the multiplexer ``MUX''). An overall dynamic phase shift $\phi_{ext}$ can be applied, as well as an individual phase shift $\phi_{det}$ on the detection clock. The user data is received over GigE by the server and is written via the driver into a memory region, reserved for direct-memory-access (DMA). The timing module reads the data via DMA from memory and uses one FIFO (TX) to buffer and transmit the data into the bus clocking region (orange). Data is read back into memory with the same DMA interface and another FIFO (RX). The auto-synchronization module generates a pulse on the trigger line and waits for its reception and a programmed number of cycles $N_w$ before it gives the start signal for the timing module to generate the data on the bus. In combination with a phase-shifted detection clock (red), the pulse round-trip number of cycles $N_{RT}$ between two boards can be measured. All control and status registers in the PL part can be accessed by the driver via the AXI Light interface and are transmitted with clock-domain-crossing (CDC \cite{CummingsCDC2008}) modules between the different clocking regions. The DMA and timing modules send interrupts (IRQ) to notify the driver of important events.}
\label{fig:logic}
\end{center}
\end{figure}

Here we give an overview on the logic used in the FPGA to generate the experiment control data on the bus and all the signals necessary for the synchronization of several boards. A simplified block diagram is shown in Fig. \ref{fig:logic}.  
The board is basically composed of two parts: the processing system (PS, top, green), consisting of a CPU on which a Linux operating system is running, and the programmable logic (PL, bottom, yellow), where our custom hardware is implemented. The two parts of the FPGA-SoC are tightly bound via interfaces and buses, enabling mutual data exchange at high speed. In such a way, the two main tasks of the board are effectively separated among the two independent parts of the FPGA-SoC system itself. While the processing system handles the communication via Ethernet with an external control computer, the logic part produces the signals on the bus. The driver mediates between both parts and coordinates the access to the external memory. The source code for programming the FPGA is written in Verilog. It is synthesized and implemented with the Vivado 2017.4 software from Xilinx running on Ubuntu 18.04 LTS, and is available online \cite{github}. Detailed information on the FPGA resources used for this application is reported in Tab. \ref{tab:resources} in Appendix \ref{sub:resources}.  

In brief, we use one general purpose I/O (GPIO) port for the reading and writing of memory mapped registers (via AMBA AXI-4 Lite interface \cite{AMBA-AXI}), and one high performance (HP) port to efficiently transfer the experiment control sequence from the memory to the PL part and vice-versa (using direct memory access DMA \cite{AXI-DMA} via an AXI stream bus). The clock frequency for the PL part, CPU and the DDR memory are set to their default values, corresponding to 50\,MHz, 650\,MHz and 525\,MHz respectively. 

The experiment control sequence (represented by thick lines in Fig. \ref{fig:logic}) is sent via Ethernet from the control computer to a TCP/IP server application running on the CPU. The server application interacts with a Linux kernel driver module \cite{github}, which writes the data into DDR memory and programs the FPGA registers using the AXI Light bus. The data are transferred via DMA from the memory into a transmit (TX) first-in-first-out (FIFO) buffer \cite{CummingsFIFO2002a,CummingsFIFO2002b} which holds a maximum of 8192 samples of 128\,bits each.  
The FIFO serves to buffer gaps in the DMA data transmission, and allows efficient transfer of data between regions using different clocks (clock domains).
In addition, we have implemented a receive DMA channel (RX), which can be used, for example, to read data from an analog input device that sends data on the bus.

In our case, the experiment control sequence consists of 64\,bits per sample: 32\,bits are used for the time-stamp, 7 address bits select which device on the bus to be updated, and 16 device specific bits define the new state of the device \footnote{\label{note96bits}An optional extended version uses 12 instead of 8 bytes per sample. This allows to have two independent buses driven by a single FPGA-SoC board with a modified buffer card.}. 
The time-stamp defines at which time the bus should be updated with the specific data and address of the corresponding device. After the bus has been updated, a pseudo-clock pulse (strobe) is generated by the FPGA on another pin of the bus, to initiate the state change of the selected device \footnote{The strobe signal is generated by the FPGA. For $\Gamma_{sample}$ = 1\,MHz it is a 500\,ns long pulse starting 240\,ns after the bus has been updated. The bus clock frequency must be at least twice the bus output rate to generate the strobe signal.}. The time-stamp is defined in units of 1/$\Gamma_{sample}$ with $\Gamma_{sample}$ being the output sampling rate of the bus, typically set to 1\,MHz or 10\,MHz.

The timing module is responsible to output the data on the bus. It first takes out one 64\,bits-wide sample from the 128\,bits of the TX FIFO, and it compares the time-stamp with an internal counter running at $\Gamma_{sample}$. When they are equal, the module outputs the 16+7 data and address bits, and it generates the previously mentioned strobe signal.
The timing module internally uses a dedicated 50\,MHz bus clock, which can be either the PL system clock (i.e. the internal oscillator of the FPGA-SoC board), or it can be generated from an external clock signal using a phase locked loop (PLL) of the FPGA-SoC. In the latter case, the frequency allowed for the external clock signal ranges from a minimum value of 10\,MHz, limited by the PLL, to a maximum of 300\,MHz, limited by the input buffer on the buffer card. A second PLL is used as a software controlled multiplexer (MUX) to switch between the two clock sources\footnote{Cascading two PLL's is not advised, but in our case, we need both for dynamic phase shifting. In addition, this allows to use an external clock input pin in a different clocking region which would be otherwise inaccessible.}. Both PLLs enable to dynamically change the phase of the generated clock signals. The auto-synchronization module, discussed in Sec. \ref{sub:auto-sync}, is using these signals to synchronize several boards. The timing module can also trigger the output of the experimental sequence, which alternatively can be started by an external hardware trigger or via software.
Finally, both the DMA TX/RX channels and the timing module communicate with the driver via interrupts. The DMA channels generate interrupts when buffers need to be updated. The timing module generates one interrupt when the experimental control sequence has been completed. Further interrupts are generated at a configurable frequency, typically 16\,Hz, and are used to update the board status in the control software. 

\subsection{Auto-synchronization}\label{sub:auto-sync}

\begin{figure}[t]
\begin{center}
\includegraphics[width=0.95\columnwidth]{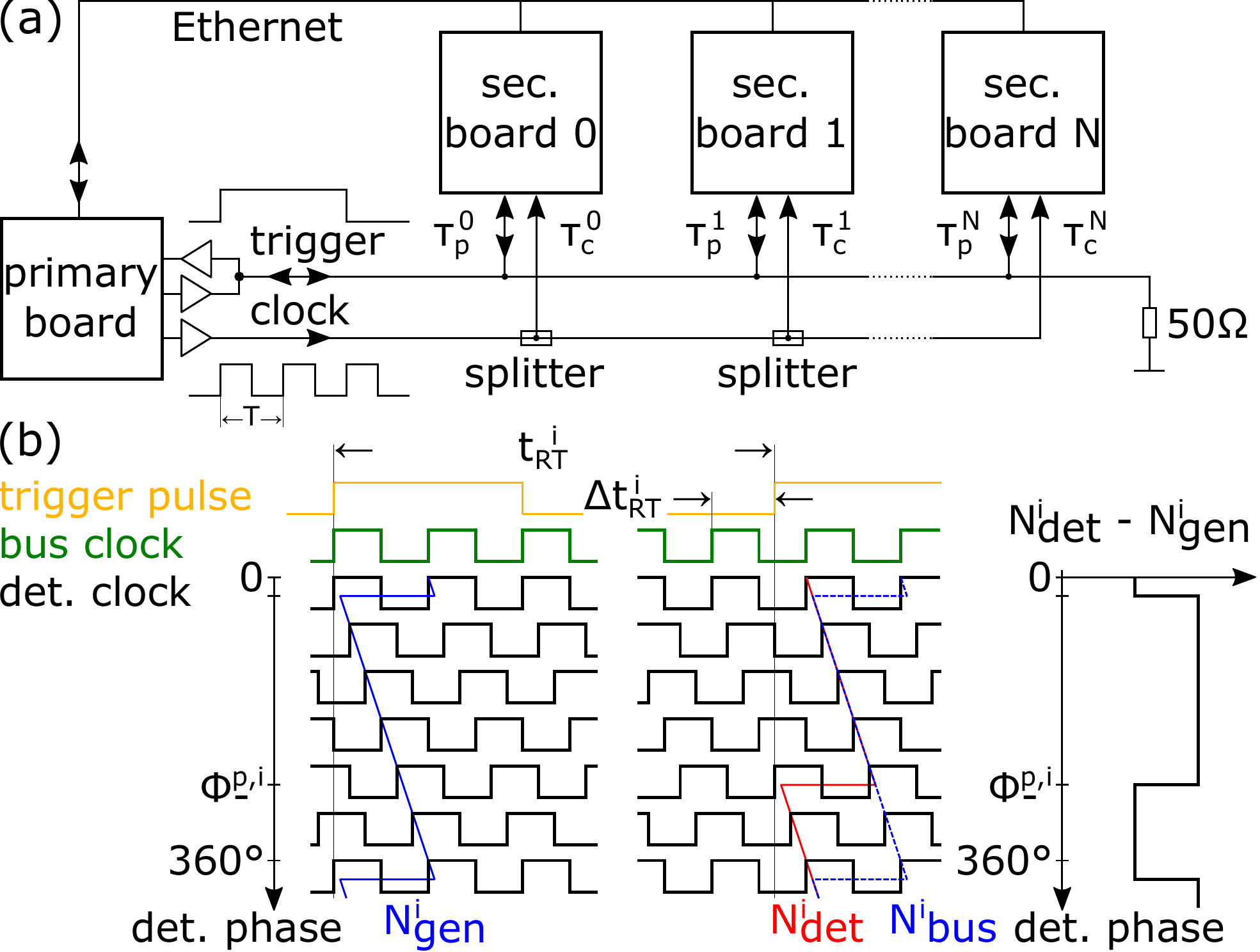}
\caption{Triggering and auto-synchronization scheme for multiple boards. a) In the simplest configuration all boards are connected with a common clock (period $T$) provided by the primary board and daisy-chained from one board to the next using splitters. Additionally, a common trigger coaxial cable directly connects all boards and is terminated by $50\,\Omega$. The primary board generates a pulse in the trigger cable which all secondary boards detect with individual delay. 
The primary and secondary boards wait until all secondary boards have received the trigger pulse and start generating output simultaneously. %
The delays between the primary and secondary boards for the clock $\tau_c^i$ and the trigger $\tau_p^i$ are indicated ($i\ \in\ 0 \ldots N$), with $N$ the number of secondary boards. 
b) The trigger delay $\tau_p^i$ of each secondary board $i$ is measured during the auto-synchronization by determining the round-trip time $t_{RT}^i$ of the pulse (orange) 
from the difference of the number of cycles from the generation of the pulse ($N_{gen}^i$, blue) and its detection ($N_{det}^i$, red). The time correction $\Delta t_{RT}^i < T$ is obtained by repeating the measurement and detecting the reflected pulse with a phase-shifted detection clock with increasing detection phase (black, seven phases shown) with respect to the bus clock (green) which is used to generate the pulse. At the phase $\phi_{-}^{p,i}$ 
the measured $N_{RT}^i$ reduces by one cycle and $\Delta t_{RT}^i$ is obtained. For board $i$ the trigger delay is calculated from $\tau_p^i = t_{RT}^i/2$. For the determination of $\Delta \tau_c^i$ a similar measurement is done on each secondary board where $\Delta t_s^i$, $N_{bus}^i$ and $\phi_{-}^{s,i}$ replace the roles of $\Delta t_{RT}^i$, $N_{gen}^i$ and $\phi_{-}^{p,i}$ in the figure. The clock delay $\Delta \tau_c^i$ at board $i$ is calculated from Eq. \eqref{eq:clock-delay-simple}. See text and Appendix \ref{sub:auto-sync-model} for more details\cite{note_cdc}
and figures \ref{fig:auto-sync-data}a and \ref{fig:auto-sync-det} for example detection signal for varying detection phase.}
\label{fig:auto-sync}
\end{center}
\end{figure}

In order to synchronize several FPGA-SoC boards, all boards need to start the experimental control sequence simultaneously and they need to use the same clock source to execute each command at the same time. The common clock can be either generated by one board, or 
provided externally. In both cases, a suitable amplification and distribution system to all boards is needed, which might introduce unknown phase shifts. Additionally, a starting (trigger) signal needs to be distributed from one board to all the others, and can accumulate an unknown delay. As discussed in the following, our scheme takes into account and corrects for both these effects. 
To compensate the delay on the start trigger signal, we adopt a scalable scheme, where one trigger line is connected with high impedance to all participating boards, see Fig. \ref{fig:auto-sync}a. The trigger line is a coaxial cable with 50\,$\Omega$ termination on both ends to avoid unwanted reflections. One board, called the primary board, receives the start signal from the control computer (or from an external hardware trigger), and generates a pulse in the trigger line which is detected by the other ``secondary'' boards. In order to compensate for the pulse propagation time between the boards, the propagation time is automatically measured in advance, such that each board can delay its execution accordingly and all boards can start at the same time.

To measure the propagation delay, the primary board instructs via Ethernet one of the secondary boards to introduce a short circuit in the trigger line using a bipolar or a field-effect transistor. Then the primary board generates a pulse in the trigger line, and it measures the round-trip time $t_{RT}^i = N_{RT}^i\,T + \Delta t_{RT}^i$ needed by the pulse to propagate to the secondary board $i$, be reflected at the short circuit, and travel back (see Fig. \ref{fig:auto-sync}b).
Here $N_{RT}^i = N_{det}^i - N_{gen}^i$ is the number of cycles between the generation ($N_{gen}^i$, blue) and the detection ($N_{det}^i$, red) of the pulse, and $\Delta t_{RT}^i < T$ is a fraction of the period $T$ of the bus clock of the primary board. 
While $N_{RT}^i$ can be measured directly, $\Delta t_{RT}^i$ cannot. This limits the resolution to the period $T$, which is 20\,ns for the chosen 50\,MHz bus clock frequency, and would not be satisfactory for bus output rates above 10\,MHz.
To measure the total delay with higher accuracy, the reflected pulse is sampled with a phase shifted replica (detection clock) of the bus clock signal. A train of trigger pulses is generated, and the phase shift of the detection clock is varied between pulses.
For a linear increase of the detection clock phase, at:
\begin{equation}\label{eq:time-corr-prim}
\phi_{-}^{p,i} = \Delta t_{RT}^i \frac{\TwoPi}{T}\ ,
\end{equation}
the measured $N_{RT}^i$ reduces by one. This change in $N_{RT}^i$ is detected, and $\Delta t_{RT}^i$ can be obtained \footnote{The actual algorithm to find the phase jump is similar to the Bisection method of finding the root of a function.}.
In principle, this method would allow one to achieve a time resolution of about 20\,ps, given the 0.3$^{\circ}$ phase resolution of the PLL at the used clock frequency. However, noise in the generation and detection of the pulse actually limits the resolution to larger values.
This measurement is repeated for each secondary board $i = 0 \ldots N$.
With the measured round-trip time $t_{RT}^i$, the propagation time of the pulse from the primary board to the $i$-th secondary board is calculated as: 
\begin{equation}\label{eq:propagation-time-simple}
\tau_p^i = t_{RT}^i/2\ .
\end{equation}
It is important to notice that in this simplified treatment we neglect all additional (but constant) delays, both internal to the FPGA and due to the electronics needed for the generation and detection of the pulse. Details of the full model accounting for these additional delays are given in Appendix \ref{sub:auto-sync-model}.

In order to achieve a perfect synchronization among all boards, the measurement of $\tau_p^i$ for each board discussed above is not sufficient, since the clocks of the secondary boards must be corrected for the delays $\tau_c^i$ 
introduced along the clock distribution line (see Fig. \ref{fig:auto-sync}a). In this case however, one needs to know only the introduced clock delay $\Delta \tau_c^i = \tau_c^i \% T$, where $\%$ is the modulus.
To this end, a second set of measurements is carried out, where the primary board generates a train of pulses similarly to the previous scheme, but the measurement is now taken on the secondary boards. Since the pulses do not need to be reflected, all the secondary boards can measure the respective clock delay simultaneously. Each secondary board determines the time $\Delta t_s^i$ between the arrival of the pulse and the previous rising bus clock edge, local to the secondary board. 
Similarly to the delay $\Delta t_{RT}^i$, here the quantity $\Delta t_s^i$ is obtained by detecting the arrival of the pulse with both the detection and the bus clock simultaneously, giving $N_{det}^i$ and $N_{bus}^i$ (blue dashed line in Fig. \ref{fig:auto-sync}b), respectively. The difference between the two signals $N_{det}^i - N_{bus}^i$ is monitored for a reduction of one cycle at the phase:
\begin{equation}\label{eq:time-corr-sec}
\phi_{-}^{s,i} = \Delta t_s^i \frac{\TwoPi}{T}\ ,
\end{equation}
and $\Delta t_s^i$ can be obtained. At the secondary board location, the calculated pulse delay with respect to the primary bus clock is $\Delta t_{RT}^i/2$ and the difference to the observed delay $\Delta t_s^i$ gives the unknown clock delay:
\begin{equation}\label{eq:clock-delay-simple}
\Delta \tau_c^i = \Delta t_{RT}^i/2 -\Delta t_s^i\ .
\end{equation}

Once $\Delta t_{RT}^i$ and $\Delta t_s^i$ are determined for each secondary board $i = 0 \ldots N$, the external clock PLL phases $\phi_{ext}^i$ of each secondary board can be set to $\phi_{ext}^i = -\Delta \tau_c^i \frac{\TwoPi}{T}$. In this way, the clocks of all secondary boards are synchronized with that of the primary one and the auto-synchronization measurement is completed and all parameters are set.
In order to simultaneously generate data on all boards, the primary board sends a pulse in the trigger line. It then waits until all secondary boards have detected the trigger pulse, i.e. it waits the largest propagation time $\tau_p^i$. Each secondary board $i$ waits $\tau_p^i$ less time than the primary board. After these waiting times, all boards synchronously start generating output of data on their bus. 

While we refer the reader to Appendix \ref{sub:auto-sync-model} for more details, we emphasize that our auto-synchronization scheme allows for the synchronization of many boards on time scales of order of nanoseconds with a relatively simple scheme and few external components. A first experimental demonstration of this scheme together with measurements of the residual synchronization timing error are presented in Sec. \ref{sec:results}.

\section{Software} \label{sec:software}

In this section we summarize the software implementation on the PS/CPU part of the SoC, on which a Linux operating system is running\footnote{Petalinux 2017.4 from Xilinx which is built on Linux kernel version 4.9 and is compiled on Ubuntu LTS 18.04.}. This is a fully featured operating system which provides system services and interfaces to external devices, and that can be configured for our specific needs. The PL part appears for the operating system like an external device, and our device driver can communicate with it via registers \cite{github}.

\subsection{Control computer software}\label{sub:control}

Many research laboratories, including ours, typically employ either Labview or LabWindows/CVI \cite{NI_Labview_CVI} as user application programs. While our setup is currently adapted to work with this software, we emphasize that any other user application can be easily implemented  on our hardware, provided that the data need to be sent via Ethernet to our TCP/IP server running on the FPGA-SoC. No additional driver nor hardware is required, and no constraints on the operating system are given for the control computer. For example, the freely-available, Python-based control software ``labscript suite'' \cite{Starkey2013} might be a viable alternative to the above mentioned commercial solutions. We provide the necessary files in Ref. \onlinecite{github} to use our FPGA-SoC board together with the suite.

In our specific case, we upgraded an existing control system based on a digital I/O card \cite{DIO64} installed on the experiment control computer, driving the bus via a 2\,m long cable and a buffer card.  The FPGA-SoC system replaces completely the former system, maintaining the compatibility with previous hardware and software. For this, a new Windows dynamic link library (DLL) has been written, which communicates via Ethernet with the FPGA-SoC while keeping the same functions of the previous I/O card.

\subsection{TCP/IP server and Linux device driver}\label{sub:driver}

We have designed a simple TCP/IP server application, running on the FPGA-SoC, which receives commands and the user data from the control computer, and which communicates with our device driver that mediates with the two FPGA-Soc parts, see Fig. \ref{fig:logic}. 

Our server application can control, via the device driver, the FPGA PL part, write the user data into reserved DMA (coherent) memory, and receive status information from it. The driver allows a user application to read back data from the PL part, wait for interrupts or for the end of the sequence. 
The driver maintains the ring buffers for the DMA transfer, and responds to the corresponding interrupts.
We have reserved 128\,MiB of memory for coherent DMA transfer. This size corresponds to $10^7$ samples and 10 seconds of contiguous data output at $\Gamma_{sample}$ = 1\,MHz. However, most applications typically do not require such a large number of samples and dense output of data. If needed, data could be uploaded via Ethernet during the experimental run as well. The reserved size
is sufficiently large to store all user data directly into coherent memory, which keeps the server and driver simple, and it avoids additional copying for repeated runs. 
A timer interrupt, generated by the PL part, and transmitted by the driver, allows the server application to send status information at regular intervals to the control computer.

\subsection{Startup script}\label{sub:linux-init}

When the board is powered up, a bootloader reads from a SD card the binary data to program the PL part and to load the required Linux image into memory, and to start the operating system. After this is completed, our startup script reads a configuration file from the SD card which contains the IP address and other information, with which it configures the Linux system and launches our TCP/IP server application. The server may either initiate the auto-synchronization procedure on startup, or wait for instructions from the control computer. A startup script and a text configuration file are used to change the configuration of the board without the need of recompiling the binary code from the sources.

\section{Measurements and results}\label{sec:results}

In this section we present and discuss measurements done on the FPGA-SoC board. For these measurements, specific code running on the FPGA-SoC system has been written, and the data has been acquired directly on the board and stored on a micro-SD card \footnote{As permanent storage medium the board uses a micro-SD (Secure Digital) card which primarily contains the Linux boot loader and boot image but can contain additional files and folders and can be used as a hard drive. The Linux image is unpacked by the bootloader in a RAM drive, but if needed it can also be expanded into a partition of the SD card. Additionally, a USB flash drive can be attached to the board for external storage.} for further analysis. Except for the verification of the synchronization error, no external measurement was needed. All the data presented in the paper is available in Ref. \onlinecite{Zenodo}.

In the first part, Sec. \ref{sub:DMA}, measurements of the DMA transmission rates are shown, defining how fast data can be transmitted from the external memory into the PL part and back. This represents a direct measure of the maximum sampling rate at which the board can contiguously output and input data.
In the second part, Sec. \ref{sub:uploading}, we present measurements on the data uploading rates over Gigabit Ethernet for both the Cora-Z7-10 and Cora-Z7-07S boards. 
This measurement confirms that Gigabit Ethernet is a good choice for experiments where a fast cycle time is required.
In the last part, Sec. \ref{sub:auto-sync-data}, we present first measurements of the proposed auto-synchronization scheme outlined in Sec. \ref{sub:auto-sync}, tested on a simple two-board configuration.  
An additional measurement presented in Appendix \ref{sub:start-stop} demonstrates the start- and stop trigger option in cycling mode.

\subsection{DMA transmission rates}\label{sub:DMA}

\begin{figure}[t]
\begin{center}
\includegraphics[width=\columnwidth]{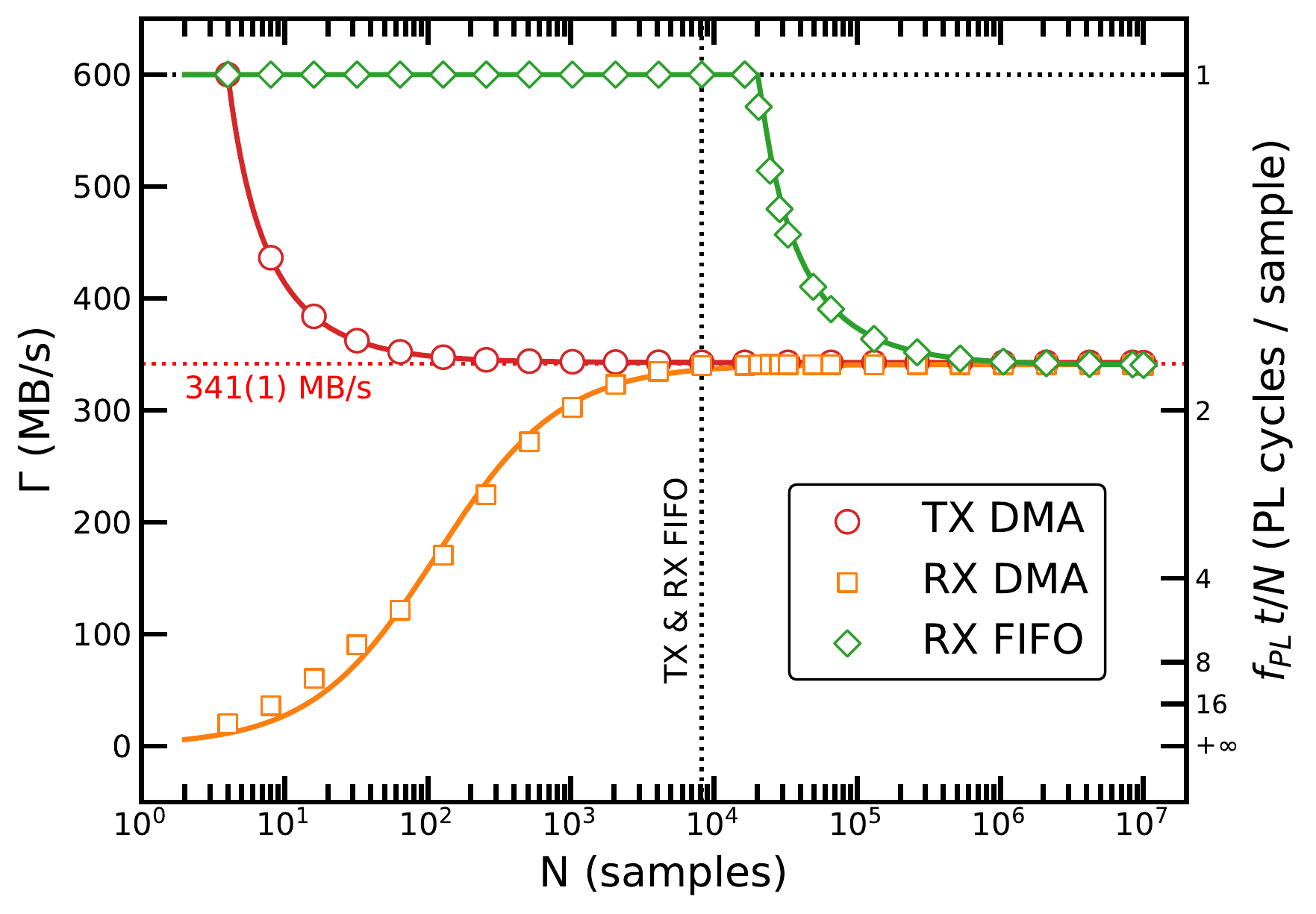}
\caption{Measurement of DMA data transmission rates of the Cora-Z7-10 board as a function of number of samples $N$. Each measurement point is the mean value of at least 20 measurements and the error bar corresponds to the standard deviation. The curves are fits to the data as explained in Appendix \ref{sub:DMA-fit}. The vertical dotted line at 8192 samples corresponds to the TX and RX FIFO buffer size. The horizontal dotted line at 600\,MB/s corresponds to 1 sample/cyle for the 50\,MHz PL clock frequency and the horizontal red dotted line is the fitted $\Gamma_{DMA}$ = 341(1)\,MB/s for large number of samples.}
\label{fig:DMA}
\end{center}
\end{figure}

In order to measure the DMA data transmission rates of the FPGA-SoC board we have temporarily added a module in the PL part which allows one to transmit data without delay in a ``loop-back'' configuration between the TX and the RX FIFO buffers (see Fig. \ref{fig:logic}), and to measure the time interval required to transmit a certain number of samples. From the measured time $t$ and the number of samples $N$ we calculate the average data rate $\Gamma$ in MB/s using: $\Gamma = \beta\,N/t$, with $\beta = 12$ bytes per sample for this measurement.
In particular, we measure three distinct rates, shown in Fig. \ref{fig:DMA} for the Cora-Z7-10 board, as a function of the number of samples $N$: the transmission rate from the memory to the PL part (TX DMA, red circles), the transmission rate from the PL part to the memory (RX DMA, orange squares) and the transmission rate through the RX FIFO (green diamonds).
Each experimental point (error bar) shown in the figure represents the mean value (standard deviation) of at least 20 repeated measurements for each $N$. 
The data are well fitted to a simple model (solid curves in Fig. \ref{fig:DMA}) that has one delay and two rates as free parameters. For details about the fitting function, and the fit results, we refer the reader to Appendix \ref{sub:DMA-fit} and Tab. \ref{tab:DMA-fit} therein.

For the measurement of the TX transmission rate (red circles in Fig. \ref{fig:DMA}) we measure the time interval from the first sample received out of the TX FIFO until the $N$-th sample is received. The first four samples are transmitted with the maximum possible rate of one sample per cycle, i.e. $\Gamma_{max} = \beta \times f_{PL}$ = 600\,MB/s (horizontal black dotted line) for the PL clock frequency of $f_{PL} =$ 50\,MHz. This is because the TX FIFO already contains three to four samples when the measurement starts (in agreement with the simulated latency of the used FIFO). As $N$ is increased, the rate reduces rapidly until it reaches a constant rate $\Gamma_{DMA}$ (horizontal red dotted line), corresponding to the transmission rate from memory to the PL part. We remark, that this characterization does not allow to measure a possible delay between the start of the DMA transmission, initiated by the CPU and the arrival of the first sample.

The second measurement (orange squares in Fig. \ref{fig:DMA}) shows the RX transmission rate obtained from the time interval between the first sample written into the RX FIFO and the RX DMA interrupt \footnote{The interrupts are generated in the PL part and are thus directly accessible during the transmission rate measurement without involving the CPU.}, which indicates that all N samples have been transmitted from the PL part to the external memory. This second rate increases for increasing $N$, from very small to the same $\Gamma_{DMA}$ as observed for the TX measurement. This initial increase is consistent with a constant delay of 202(8) PL cycles, required for the RX DMA channel to start or finish the transmission. This delay is larger than expected \footnote{On the TX DMA side we observe a delay of about 30 cycles between the arrival of the last data out of the FIFO and the TX interrupt.}, and it points to a significant latency in the RX channel. Nonetheless, the large RX FIFO can easily compensate for such a latency.

The third measurement, shown in Fig. \ref{fig:DMA} as green diamonds, was taken simultaneously with the RX transmission rate, and it shows the data rate through the RX FIFO: namely, the rate obtained from the time N samples need to pass through the RX FIFO during active RX transmission. As long as the RX FIFO is not full, one sample per cycle is transmitted, corresponding to $\Gamma_{max}$. When the RX FIFO becomes full with $N_{FIFO} = 8192$ samples (dotted vertical line in Fig. \ref{fig:DMA}), the rate reduces to the RX and TX data transmission rate $\Gamma_{DMA}$. Since the RX FIFO is simultaneously loaded with $\Gamma_{max}$, and unloaded with $\Gamma_{DMA}$, we expect this rate to drop once the number of transmitted samples reaches $N_{FIFO} \frac{\Gamma_{max}}{\Gamma_{max}-\Gamma_{DMA}} \approx 19\times 10^3$ samples, a value close to the observed one of $20(1)\times10^3$ samples. 

All three measurements give for large number of samples a consistent DMA transmission rate of $\Gamma_{DMA} = 341(1) MB/s$ (averaged over all measurements).
This rate deviates with the specified rates from Xilinx \cite{AXI-DMA} for the default settings. In particular, the TX rate is lower while the RX rate is higher than specified. However, their measured sum is 684(2)\,MB/s, which is only 2\% lower than the value expected from the specification of 700\,MB/s. 
Although the exact reason for this discrepancy is not clear (the ratio between the TX and RX rates can be adjusted \cite{Zynq-SDK-performance, SoC-performance}), the observed overall performance allows us to conclude that our DMA transmission rates are indeed close to the maximum possible ones for a single HP port. Finally, from the measured  DMA transmission rate we can also directly deduce the maximum contiguous bus data rate of $\Gamma_{DMA}/\beta$ = 30 - 40\,MHz~\footnote{The measured $\Gamma_{DMA}$ corresponds to a maximum $\Gamma_{sample}$ of 42\,MHz (28\,MHz) for the 8 (12) bytes per sample versions. The given rates apply independently for data output and input on the bus and for simultaneous output and input (if the bus supports).}.
 
We note that, the FPGA-SoC has 4 HP ports, and in our design there should be enough free resources to use at least an additional one to increase the DMA rate even further, if higher bus rates are needed. Short ``bursts'' of data output (input) of up to 8192 samples at higher frequencies are already possible with the present setup as long as there is sufficient time before the ``burst'' to fill (empty) the TX (RX) FIFO and the rate afterwards is slow enough to prevent the TX (RX) FIFO from becoming empty (full). Although not shown here, we have performed the same measurement for the Cora-Z7-07S board, finding no significant deviation from the results presented in Fig. \ref{fig:DMA}.

\subsection{Ethernet uploading rates}\label{sub:uploading}

\begin{figure}[t]
\begin{center}
\includegraphics[width=\columnwidth]{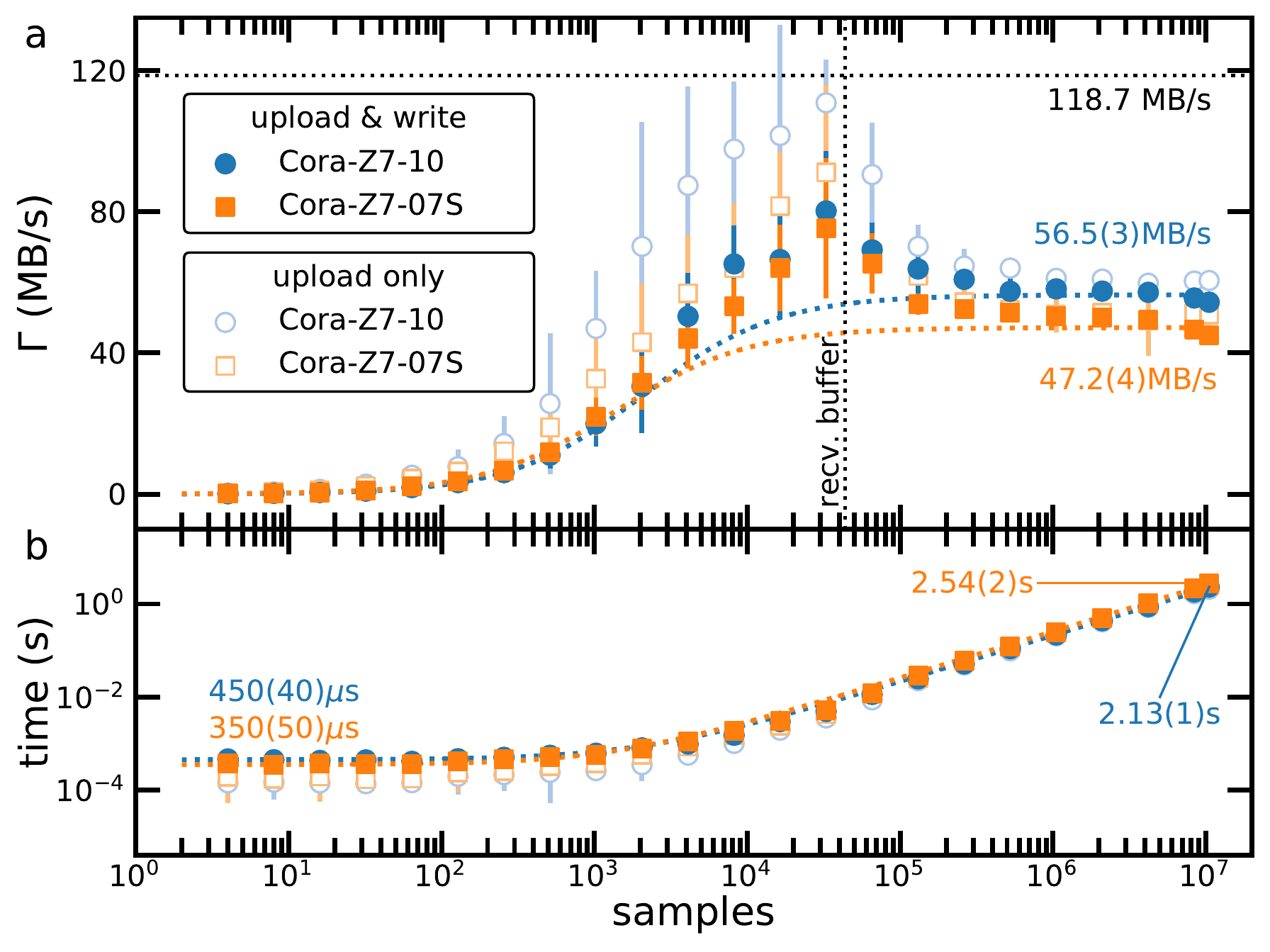}\\
\caption{a) Measured rates for uploading and writing to reserved DMA memory (solid symbols) and uploading only (open symbols) as a function of number of samples $N$ for the Cora-Z7-10 (blue circles) and Cora-C7-07S board (red squares). The horizontal dotted line indicates the theoretical maximum rate of 118.7\,MB/s for Gigabit Ethernet \cite{GigE} and the vertical dotted line indicates the size of the receive buffer of the server. The numbers are the measured uploading and writing rates for $10^7$ samples. Each data point is the mean of at least 15 measurements and the error represents the standard deviation. The dotted curves are fits with Eq. \eqref{eq:fit-model} with a delay and single rate and the fit results are summarized in Tab. \ref{tab:DMA-fit}. b) Same data as in panel a but time for uploading or uploading and writing to memory is shown. Numbers give the \chg{fitted} time needed for uploading and writing to DMA memory for 4 samples and \chg{$10^7$} samples for the Cora-Z7-10 (blue) and Cora-Z7-07S (orange) boards.}
\label{fig:uploading}
\end{center}
\end{figure}

The uploading rate from the control computer to the FPGA-SoC board over Gigabit Ethernet is another measure of the performance of our system. It can be a limitation for experiments where short cycle times are need, like experiments with optical tweezers \cite{Lukin2016} or with ions \cite{Ott2017}.

Fig. \ref{fig:uploading}a shows the uploading rate measured for the Cora-Z7-10 (solid blue circle) and Cora-Z7-07S (solid orange square) board. This measurement includes the total time of uploading and writing into reserved DMA memory. For each board the fastest strategy is used depending if a dual-core CPU is present (Cora-Z7-10) or only a single-core CPU (Cora-Z7-07S): for the dual-core CPU the server uses one thread to receive the uploaded data and a second thread to write the data into reserved DMA memory in parallel. For the single-core CPU it is fastest to immediately write the uploaded data into reserved DMA memory using a single thread\footnote{The change in the rate between using a single or two threads on both boards is only about 10\%.}. Fig. \ref{fig:uploading}b shows the corresponding times for the same data as in Fig. \ref{fig:uploading}a.

The rates are calculated from $\Gamma = N \beta / ( t_{tot} - t_{ACK} - t_{RT}^{net}/2 )$ where $N$ is the number of transmitted samples and $\beta$ = 12 bytes per samples used for the measurement. The time $t_{tot}$ is when uploading and writing to memory is finished, and $t_{ACK}$ is the time when the server acknowledged to receive the data from the user application. The network round-trip time $t_{RT}^{net}$ is obtained during each individual measurement as the time from the acknowledge of the server ($t_{ACK}$) until the arrival of the first data at the server. We take half of $t_{RT}^{net}$ under the assumption that sending and receiving involves the same delays, which is not necessarily the case. For each datapoint we have taken at least 15 measurements and plot the mean value and standard deviation (error bar). 

For small number of samples the observed uploading rate is small. This can be interpreted as a fixed delay (of order of a few 100$\,\mu$s, see Fig. \ref{fig:uploading}b), which the user application or the server needs to start sending or receiving the data. For increasing number of samples, this delay becomes less important and the rate reaches a peak of about 70 - 80\,MB/s at $32 \times 10^3$ samples (vertical dotted line) and decreases for number of samples beyond this. At $10^7$ samples the uploading and writing rate is 56.5(3)\,MB/s (47.2(4)\,MB/s) for the Cora-Z7-10 (Cora-Z7-07S) board, which corresponds to a time of \chg{2.13(1)\,s (2.54(2)\,s)}. This time is even faster than the typical calculation time the user application needs (about 7\,s with labscript-suite) to generate this number of samples.

The peak in the rate is correlated with the receive buffer size (512\,kiB) of the server. If chosen too small the decrease in the rate at higher $N$ becomes much worse. This indicates that the overhead in handling large lists of small buffers can become significant. In this respect the Cora-Z7-10 board performs slightly better than the Cora-Z7-07S board, which is limited by a single-core CPU.
 
For comparison, we present another measurement where only data are uploaded, but no writing to the reserved DMA memory is done. The resulting rates for the Cora-Z7-10 (open blue circle) and Cora-Z7-07S (open orange square) board are shown in Fig. \ref{fig:uploading}a and b. For the calculation of the rate, $t_{tot}$ is now the time until all data is uploaded without writing to reserved DMA memory. For the Cora-Z7-10 board the peak uploading rate reaches about 110\,MB/s which is very close to the theoretical maximum of 118.7\,MB/s for Gigabit Ethernet \cite{GigE}. The Cora-Z7-07S board is with about 90\,MB/s only slightly slower. In this measurement the CPU is still copying data into temporary buffers which explains the difference of the boards, and the observed decrease of the rate after the peak. 

With Eq. \eqref{eq:fit-model} in Appendix \ref{sub:DMA-fit} we fit the measurements with a delay time and a single transmission rate (dotted curves in Fig. \ref{fig:uploading}). We use the standard deviation of each data point to get more weight on the large number of samples with less noise. See Tab. \ref{tab:DMA-fit} for the fit results. The numbers in the figure are the fitted rates and times for both boards when uploading and writing 10.5 $\times 10^6$ samples to reserved DMA memory.

The observed fast uploading and writing rates confirm that the FPGA-SoC board is indeed the right choice for applications where fast cycle times are requested. 

\subsection{Auto-synchronization}\label{sub:auto-sync-data}

\begin{figure}[t]
\begin{center}
\includegraphics[width=\columnwidth]{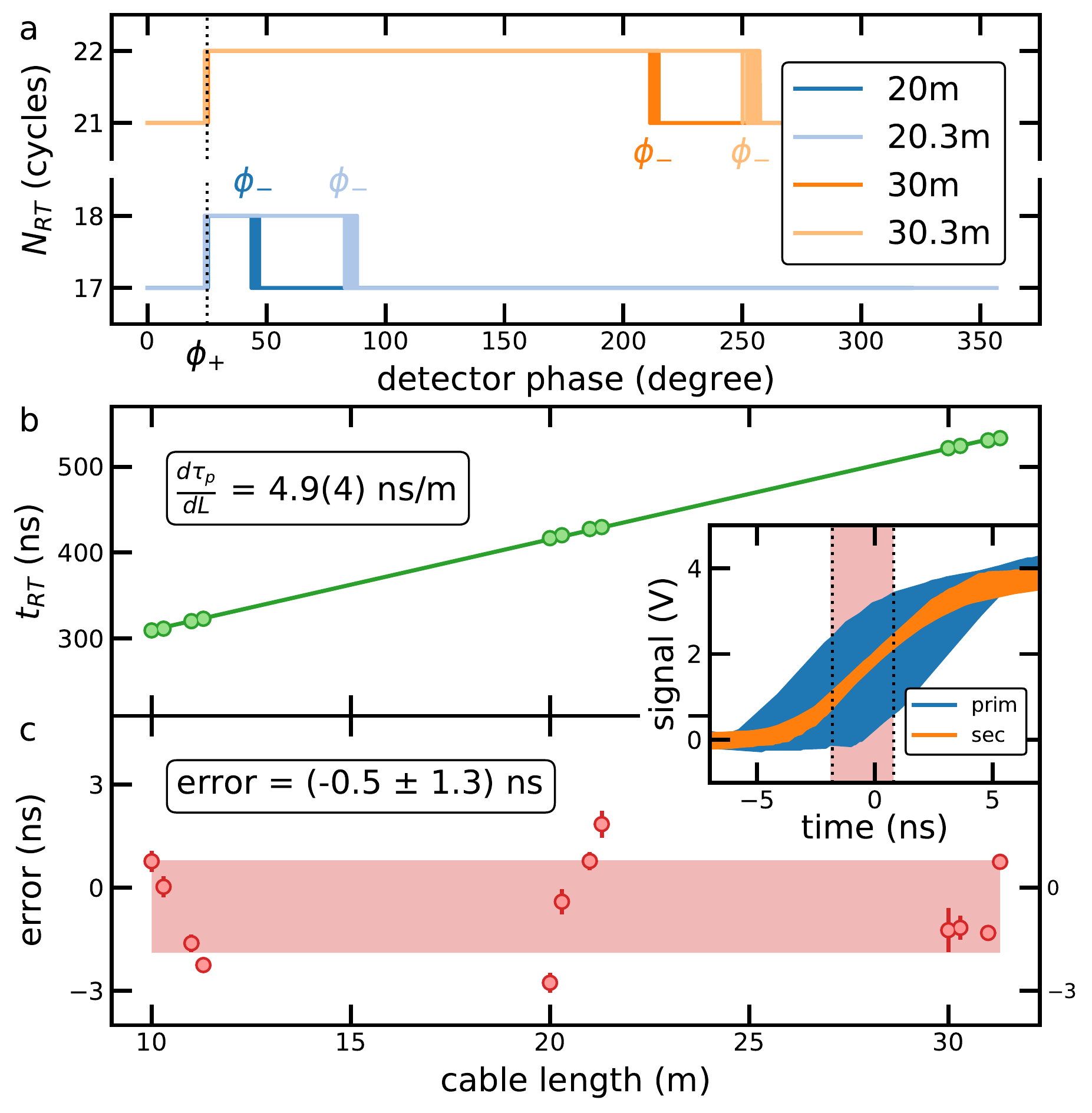}\\
\caption{Auto-synchronization result for two boards at different trigger cable lengths. a) Round-trip cycle time $N_{RT}$ for the reflected pulse leading edge vs. detector phase shows jumps of one cycle at specific phases ($\phi_{+}$ and $\phi_{-}$, see Sec. \ref{sub:auto-sync} and Appendix \ref{sub:auto-sync-model} for details). Data is shown for selected cable lengths. 
b) Pulse round-trip time $t_{RT}$ calculated with Eq. \eqref{eq:propagation-time} for the trailing edge of the pulse for 12 cable lengths. The slope of the linear fit gives a propagation delay per unit cable length of $\frac{d\tau_p}{d L}$ = 4.9(4)\,ns/m, when averaged over leading and trailing edges of the pulse. c) Synchronization error as a function of cable length. Each point and error bar is the mean and standard deviation of five repetitions with external clock phase 0, 90, 180 and 270$^{\circ}$. The red shaded area represents the 68\% confidence interval of the average error over all data giving (-0.5 $\pm$ 1.3)\,ns. 
The insert shows all signal traces of the primary (blue) and secondary board (red) used to measure the synchronization error.}
\label{fig:auto-sync-data}
\end{center}
\end{figure}

Here we present the first realization of the auto-synchronization scheme proposed in Sec. \ref{sub:auto-sync}. In particular, first tests have been done utilizing two boards connected with different trigger cable lengths and using different external clock phases. 
Without loss of generality, we present the synchronization of the two boards that are directly connected with the trigger line, terminated with 50\,$\Omega$ on the primary board side and switchable on the secondary board side from 50\,$\Omega$ to high impedance to reflect the pulse.
In the following we omit the index $i = 0$ since here only one secondary board is used.
For details on the theoretical analysis and the measurement of the secondary board external clock PLL phase we refer the reader to Appendix \ref{sub:auto-sync-model} and \ref{sub:auto-sync-phase}.

On the primary board we measure the round-trip cycle time $N_{RT}$ of the reflected pulse, and the phase $\phi_{-}^p$ at which $N_{RT}$ is reduced by one, see Fig. \ref{fig:auto-sync-data}a for different lengths of the trigger coaxial cable\footnote{For cable lengths $< 3$\,m the actual setup cannot detect the round-trip time since the reflected pulse is too close to the generated one. However, this situation is automatically detected and with the proposed scheme and further technical improvements shorter cables should be detectable.}. Combining both measured values of $N_{RT}$ and $\phi_{-}^p$ we obtain, from Eq. \eqref{eq:propagation-time} in Appendix \ref{sub:auto-sync-model}, the round-trip time $t_{RT}$ shown in Fig. \ref{fig:auto-sync-data}b. 
From a linear fit to the data (green line) we obtain the propagation delay per unit of cable length $L$ of $\frac{d\tau_p}{d L}$ = 4.9(4)\,ns/m, when averaged over leading and trailing edges of the pulse. This value is consistent with the expected one \cite{RG58}. 

Based on a similar measurement protocol\footnote{For the measurement on the secondary board the pulse is not reflected to avoid interference of the incoming with the reflected pulse. However, we have not observed a difference in the measurement result.}, the secondary board determines the phase $\phi_-^{s}$ of the negative jump in $N_{det}-N_{bus}$ for the received pulse. The local clock of the second board is locked to the external clock provided by the primary one, where a short (ca. 20\,cm long) cable is employed to ensure no additional phase shifts. To simulate different delays $\Delta \tau_c$ of the external clock, four different auto-synchronization measurements are performed, where the external clock PLL phase of the secondary board is set to 0, 90, 180 or 270$^{\circ}$, corresponding to $\Delta \tau_c$ = 0, 5, 10 or 15\,ns respectively. 

The resulting synchronization error is verified in a final measurement for each cable length and $\Delta \tau_c$ after the auto-synchronization is finished, see Fig. \ref{fig:auto-sync-data}c. For this measurement, 
the resulting phase $\phi_{ext}$, obtained from Eq. \eqref{eq:clock-phase} in Appendix \ref{sub:auto-sync-model}, is added to the previously set external PLL clock phase $\Delta \tau_c \frac{\TwoPi}{T}$, which, for perfect synchronization, should be compensated by $\phi_{ext}$.
Then the primary board generates a trigger pulse and waits $N_w^{prim} = \tau_p \sslash T$ cycles (see Eq. \eqref{eq:detection-time} and \eqref{eq:waiting-time} in Appendix \ref{sub:auto-sync-model}; the symbol $\sslash$ represents integer division), before it starts generating data on the bus. The secondary board starts generating data on the bus as soon as the trigger signal is detected.
The synchronization error corresponds to the difference between the times at which secondary and primary boards start generating data on their own buses. The corresponding traces are recorded with an oscilloscope, see the inset of Fig. \ref{fig:auto-sync-data}, and are fitted with a sigmoid function to obtain the synchronization error. See Appendix \ref{sub:trace-fit} for further details. 
In Fig. \ref{fig:auto-sync-data}c each data point (error bar) represents the mean (standard deviation) of the synchronization error, measured at least five times for each of the four external clock phases ($\Delta \tau_c$). Averaging over all cable lengths, we obtain a synchronization error of (-0.5 $\pm$ 1.3)\,ns (red shaded area in Fig. \ref{fig:auto-sync-data}c) which is much smaller than the 25\,ns time resolution for the maximum possible bus output rate of 40\,MHz of the board.

Finally we remark that, although the basic principle of our auto-synchronization scheme is very simple, being based on a round-trip time measurement, the details can be involved. Developing such a scheme on a FPGA-only platform is feasible, but it might be challenging and time-consuming. In turn, our FPGA-SoC board allows one to implement a simple pulse generation and detection in hardware, but to analyze the data and calculate the ideal settings to minimize the error, via the CPU, by software. In this way, the system could be quickly developed, errors corrected and the formulas implemented in software with no need to change the hardware every time. We believe that, the auto-synchronization is not only a useful feature, but it is also a perfect example of the flexibility which the FPGA-SoC approach offers. 

\section{conclusions and outlook}\label{sec:conclusions}

In conclusion, we have successfully implemented a versatile experimental control system based on a commercial, low-cost, and stand-alone FPGA-SoC board. %
We have demonstrated that the board can sustain bus output and input rates of up to 40\,MHz and we have 
shown how the board can automatically synchronize with a timing error approaching 1\,ns.
Furthermore, we have proven the extreme flexibility, easy Ethernet connectivity, and computational power of the FPGA-SoC system, showing several examples in which the operating system, running on the board itself, is used not only to control the FPGA hardware, but also for data acquisition and analysis. Finally, we stress that no specific device driver or proprietary software, or operating system is needed to use our device, and that the whole source code to program the FPGA-SoC is freely available \cite{github}. 
Although not discussed in the present work, our system can be easily extended to include the control of additional devices through the on-board USB host controller \cite{USBTMC}, or via adapter with the older GPIB standard \cite{GPIB}, widespread in many laboratories, or to directly read data with analog-to-digital converters (ADC).
We also emphasize that our design is stand-alone and lightweight, and the power consumption of less than 2\,W, makes it compatible for the operation in remote locations, and even for experiments in space \cite{ClockSpace2018,BECinspace2020,InterferometerSpace2021}.
We believe that the auto-synchronization feature, devised and implemented in this work, will also help several experimental setups on ground with growing complexity: for instance, setups which must bridge large distances to challenge relativity \cite{Hensen2015}, to detect gravitational waves with large-scale atom interferometers \cite{Kasevich2013,Bouyer2018}, and to measure difference of gravitational red-shift between two separated atomic lattice clocks \cite{Katori2020}. 
Finally, our architecture, thanks to the rich features and flexibility offered by the new FPGA-SoC board, may find application in various research fields, extending well beyond our original purpose of controlling AMO physics experiments.

\begin{acknowledgments}
We thank Jacopo Catani for fruitful discussions, borrowing equipment and careful reading of the manuscript, Roberto Concas and Fabio Corti for machining and soldering a prototype buffer card, Giacomo Mazzamuto for help with github, and all members of the Quantum Gases Group at LENS, in particular Leonardo Fallani and Daniele Tusi and the Yb team for testing the boards in their experiment. This work was supported by the ERC through grant No. 637738 PoLiChroM and by the Italian MIUR through the FARE grant No. R168HMHFYM P-HELiCS. N.P.  acknowledges  support  from  European  Research  Council,  Grant No. 772126 (TICTOCGRAV).
\end{acknowledgments}

The authors declare that they have no competing interests.

\section*{Data Availability Statement}

\begin{center}
\renewcommand\arraystretch{1.2}
\begin{tabular}{| >{\raggedright\arraybackslash}p{0.3\linewidth} | >{\raggedright\arraybackslash}p{0.65\linewidth} |}
\hline
\textbf{AVAILABILITY OF DATA} & \textbf{STATEMENT OF DATA AVAILABILITY}\\  
\hline
Data openly available in a public repository that issues datasets with DOIs
&
The data that support the findings of this study are openly available at https://doi.org/10.5281/zenodo.4893285
\\\hline
\end{tabular}
\end{center}

\appendix

\section{Auto-synchronization}

In Sec. \ref{sub:auto-sync-model} we present the full model of the auto-synchronization scheme outlined in Sec. \ref{sub:auto-sync} and in Sec. \ref{sub:auto-sync-phase} we show additional data for the first implementation presented in Sec. \ref{sub:auto-sync-data}. In Sec. \ref{sub:trace-fit} the fitting function is presented which is used to obtain the synchronization error shown in Fig. \ref{fig:auto-sync-data}c in Sec. \ref{sub:auto-sync-data}. In Sec. \ref{sub:auto-sync-det} sample detector signals are shown.

\subsection{Theoretical Model}\label{sub:auto-sync-model}

\begin{figure}[t]
\begin{center}
\includegraphics[width=\columnwidth]{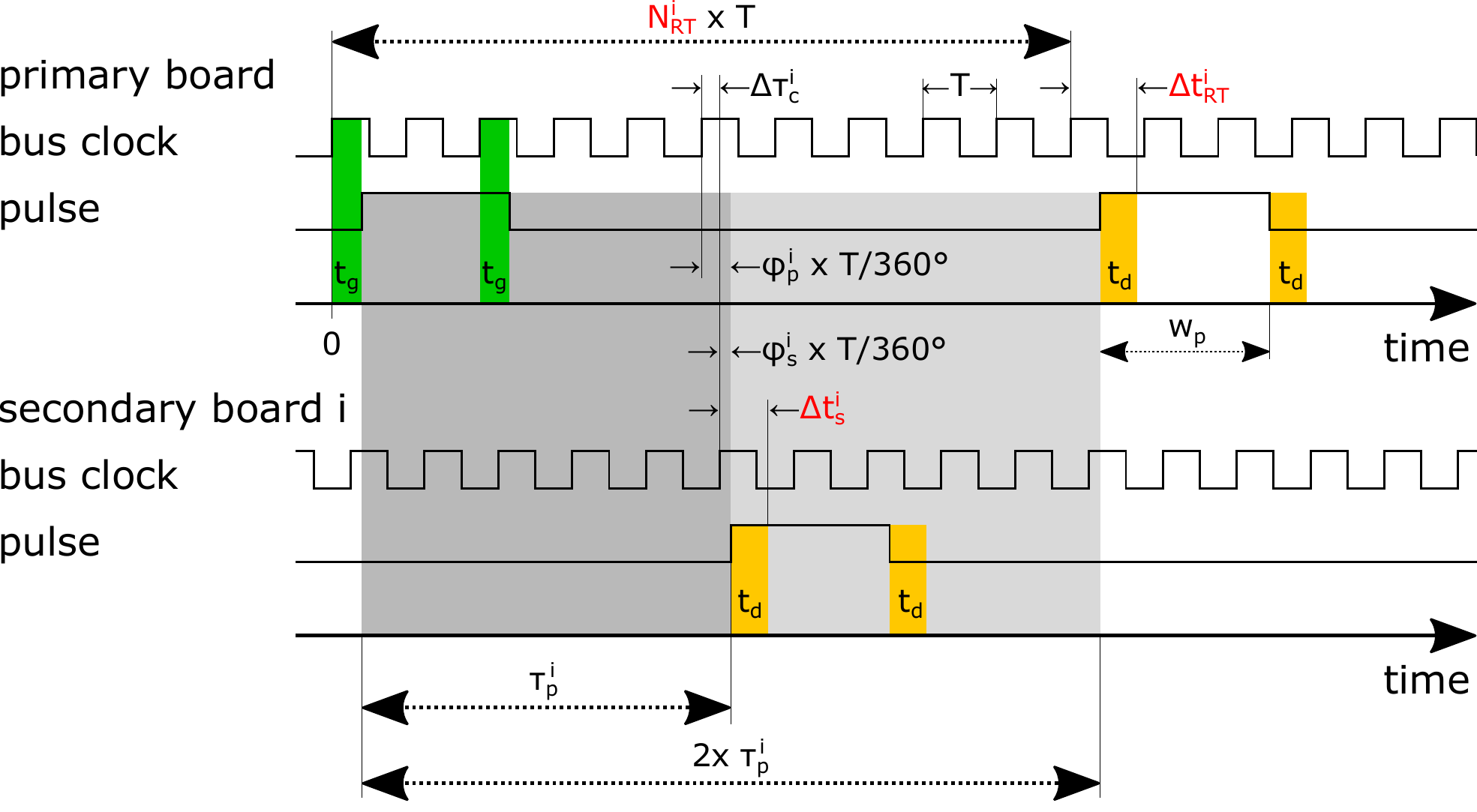}\\
\caption{Graphical representation of main quantities (delays and phases) involved in the auto-synchronization scheme. Upper part: the primary board generates the pulse and waits until detection of the reflected signal after a propagation time of $2 \times \tau_p^i$ (light gray). Delays involving the generation ($t_g$, green) and the detection ($t_d$, orange) of the pulse have to be added for the calculation of the total round-trip time $t_{RT}^i = N_{RT}^i\,T + \Delta t_{RT}^i$, with $T$ the clock cycle time. Lower part: the secondary board $i$ detects the pulse after the propagation time $\tau_p^i$ (dark gray) and it is assumed the same delays as for the primary board. The delay of the local clock of the secondary board with respect to the primary board is $\Delta \tau_c^i$ and can be calculated from the difference of $\varphi_p^i - \varphi_s^i$. The measured quantities $N_{RT}^i$, $\Delta t_{RT}^i$ and $\Delta t_s^i$ are indicated in red. The width of the pulse $w_p$ is changing during the propagation due to dispersion, and affects the measurement if this involves both leading and trailing edges of the pulse (not shown here).}
\label{fig:auto-sync-timing}
\end{center}
\end{figure}

A graphical representation of all the quantities and delays involved in the synchronization scheme is presented in Fig. \ref{fig:auto-sync-timing} for the primary and secondary boards. The measurement on the primary board gives for each secondary board $i$ the round-trip number of cycles $N_{RT}^i = N_{det}^i-N_{gen}^i$ and the negative jump in $N_{RT}^i$ gives $\Delta t_{RT}^i$ from Eq. \ref{eq:time-corr-prim}. On the secondary board the time $\Delta t_s^i$ is measured from the negative jump in $N_{det}^i-N_{bus}^i$ using Eq. \ref{eq:time-corr-sec}. From these quantities the waiting number of cycles $N_w^i$ and the external clock phase $\phi_{ext}^i$ and the detector phase $\phi_{det}^i$ (see Fig. \ref{fig:logic}) are calculated as described below. 

The model uses a set of constants which are summarized in Tab. \ref{tab:constants}. They have been determined from several calibration measurements, or have been chosen for best performance, as described below. The PL system clock is 50\,MHz for this measurement, but it should affect only $t_g + t_d$ (see below) through the fixed number of clock cycles used for the CDCs. After these parameters have been determined, they can be applied for all boards and should not need to be changed as long as the boards are the same and the setup (hardware and software) is not changed.  

\begin{table}[tb]
\begin{minipage}{\columnwidth}
\begin{tabular}{lll}
name &value &remark\\
\hline
$t_g + t_d$ &205(1)\,ns &offset from linear fit Fig. \ref{fig:auto-sync-data}b\footnotemark[1]\footnotemark[2]\\
$t_d$ &-2(1)\,ns &offset from linear fit Fig. \ref{fig:auto-sync-phase} at $\Delta \tau_c = 0$\,ns\footnotemark[1]\\
$\phi_{+}$ &25(1)$^{\circ}$ &measured\footnotemark[2]\\
$\varphi_p^{crit}$ &180(20)$^{\circ}$ &measured\footnotemark[3]\\
$\phi_{0}$ &20$^{\circ}$ &fine-adjusted manually to minimize the error\\
$N_0$ &3 &adjusted manually to minimize the error\\
$\varphi_m$ &90$^{\circ}$ &chosen\\
$\varphi_{add}$ &70$^{\circ}$ &chosen\\
$\Delta \varphi_p^{crit}$ &20$^{\circ}$ &chosen\\
$\delta \varphi_p^{crit}$ &30$^{\circ}$ &chosen\\
\hline
\end{tabular}
\caption{Used constants for the auto-synchronization. The measured standard deviation is given in brackets.}
\label{tab:constants}
\footnotetext[1]{Obtained from earlier measurements.}
\footnotetext[2]{At $f_{PL}$ = 50\,MHz.}
\footnotetext[3]{Error is smaller than $\Delta \varphi_p^{crit}$ but was not systematically measured.}
\end{minipage}
\end{table}

Taking into account the generation time $t_g$ (green) and the detection time $t_d$ (orange) of the pulse, the round-trip time $t_{RT}^i$ and propagation time $\tau_p^i$ between the primary and the secondary board is obtained from:
\begin{equation}\label{eq:propagation-time}
\begin{gathered}
t_{RT}^i = \left\{
\begin{array}{c}
N_{RT}^i \\
N_{RT}^i + 1
\end{array}
\right\} \,T + \Delta t_{RT}^i 
 \quad \text{for}\quad  \left\{ 
\begin{array}{c}
\phi_{-}^{p,i} > \phi_{+} \\
\phi_{-}^{p,i} < \phi_{+}
\end{array}
\right. \\
\tau_p^i = \left( t_{RT}^i - t_g - t_d \right) / 2\ .
\end{gathered}
\end{equation}
This is the full relation in comparison to Eq. \ref{eq:propagation-time-simple} in Sec. \ref{sub:auto-sync}. 
At the phase $\phi_{+}$ of the detector clock, the measured $N_{RT}$ increments by one cycle. $\phi_{+}$ is at a small and positive detector phase, because the signal for $N_{gen}$ (blue solid line in Fig. \ref{fig:auto-sync}b) has to be transmitted from the bus clock to the detection clock and for too small delay between the clock edges the signal is transmitted one cycle later. For detector phases above $\phi_{+}$ the signal can be transmitted within the same clock cycle\footnote{If $\phi_{-} \approx \phi_{+}$ the measurement is not reliable due to its sensitivity to noise.}.
Therefore, one cycle has to be added to $t_{RT}^i$, when the measured \mbox{$\phi_{-}^{p,i} < \phi_{+}$}. 
This happens regardless of the additional clock-domain-crossing stage (CDC, see Fig. \ref{fig:logic}, avoided in Fig. \ref{fig:auto-sync}b for clarity), which is needed for the transmission of the signal for $N_{gen}$ from the bus clock to the detection clock.
The sum $t_g + t_d$, used for the calculation of the propagation time $\tau_p^i$, is the experimentally obtained offset of the linear fit of the round-trip time vs. cable length (see green line in Fig. \ref{fig:auto-sync-data}b).

From the propagation time $\tau_p^i$ the pulse phase $\varphi_p^i$ can be calculated:
\begin{equation}\label{eq:pulse-phase-prim}
\begin{aligned}
\varphi_p^i &= \left( \left(\tau_p^i + t_g \right) \% T \right) \frac{\TwoPi}{T} \\
&= \left(\frac{\Delta t_{RT}^i + t_g - t_d + (t_{RT}^i \sslash T)\,T}{2} \% T\right)\frac{\TwoPi}{T} \ .
\end{aligned}
\end{equation}
The symbols $\%$ and $\sslash$ represent modulo and integer division, respectively. The factor $(t_{RT}^i \sslash T)\,T$ adds $T/2$ to $\varphi_p^i$ when $\phi_{-}^{p,i} < \phi_{+}$. $\varphi_p^i$ 
is the expected phase of the pulse which the secondary board would measure for $\Delta \tau_c^i = 0$. The actual pulse phase which the secondary board obtains is:
\begin{equation}\label{eq:pulse-phase-sec}
\varphi_s^i = \left(\Delta t_{s}^i - t_d\right) \frac{\TwoPi}{T} \ ,
\end{equation}
where we assume that the detection delay $t_d$ is the same as for the primary board. The difference between the primary and secondary pulse phase is a measure of the secondary clock delay $\Delta \tau_c^i$. This is used to set the external clock phase $\phi_{ext}^i$ of the secondary board:
\begin{equation}\label{eq:clock-phase}
\phi_{ext}^i = -\Delta \tau_c^i \frac{\TwoPi}{T} = \varphi_s^i - \varphi_p^i - \phi_0 + \xi(\varphi_p^i)\ .
\end{equation}
This is the full relation corresponding to Eq. \ref{eq:clock-delay-simple} in Sec. \ref{sub:auto-sync}. The additional phase factor $\phi_0$ is manually adjusted to minimize the synchronization error. This corrects an eventual mismatch in $t_d$ between the primary and secondary board and corrects for our choice to measure $\varphi_p^i$ on the trailing edge and $\varphi_s^i$ on the leading edge of the pulse\footnote{This choice was motivated to have similar $\frac{d\tau_p}{d L}$ for the measurements of the primary and secondary board. The average in $\frac{d\tau_p}{d L}$ for the leading and trailing edge of the pulse is the same for both boards, but the primary board shows a larger discrepancy between the values obtained for the two edges. The difference is caused by the dispersion of the pulse. $\phi_0$ corrects the phase shift introduced by half of the pulse width $w_p/2$ (see Fig. \ref{fig:auto-sync-timing}) but does not correct for the changing width along the path.}.
When $\varphi_p^i$ happens to be close to the critical phase $\varphi_p^{crit}$, the resulting synchronization error shows random jumps by $T$ in either positive or negative direction\footnote{The value of $\varphi_p^{crit}$ (see Tab. \ref{tab:constants}) has been determined experimentally, but there might be a dependence with our choice of parameters. Its exact origin has not been investigated.}. The security phase $\xi(\varphi_p^i)$ is introduced to avoid this region which we define as $\pm \Delta\varphi_p^{crit}$ around $\varphi_p^{crit}$. $\xi(\varphi_p^i)$ is nonzero only if $\varphi_p^i$ is inside this region and adds in this case $\pm \delta\varphi_p^{crit}$ to $\phi_{ext}$ according to:
\begin{equation}\label{eq:security-phase}
\begin{aligned}
\xi_0^i &= -sign(\varphi_p^i - \varphi_p^{crit})\times \delta\varphi_p^{crit}\\
\xi(\varphi_p^i) &= \left\{ 
\begin{array}{cl}
\xi_0^i &\text{for} \ |\varphi_p^i - \varphi_p^{crit}| < \Delta\varphi_p^{crit}\\
0 &\text{otherwise}
\end{array}
\right.\ .
\end{aligned}
\end{equation}
The function $sign(x)$ gives $\pm 1$ depending on the sign of $x$. When $\xi(\varphi_p^i)$ is nonzero, the synchronization error increases by about $\delta\varphi_p^{crit} \frac{T}{\TwoPi} \approx$ 1.7\,ns, but avoids uncontrollable outliers. The data points at 20\,m and 31.3\,m in Fig. \ref{fig:auto-sync-data}c and \ref{fig:auto-sync-data}d represent such cases where the measured $\varphi_{p}$ is about $\pm 15^{\circ}$ near $\varphi_p^{crit}$ (see green shaded region in Fig. \ref{fig:auto-sync-phase}). 
Note, that this correction depends only on the measured $\varphi_{p}^i$ and is automatically applied by the boards. For applications where the added synchronization error is unacceptable, the board can give a warning to the user and a slightly shorter or longer trigger cable might be used.

The detection clock phase $\phi_{det}^i$ is used not only during the auto-synchronization measurement, but also afterwards to detect the pulse on the secondary boards. It does not directly influence the synchronization error, but it is set such that the detection of the trigger pulse happens neither close to the rising or falling edges of the pulse, nor to the rising edge of the bus clock. This ensures reliable timing but might require one additional cycle to wait. $\phi_{det}^i$ is set at least $\varphi_{add}$ after the arrival of the pulse: 
\begin{equation}\label{eq:detector-phase}
\begin{gathered}
\varphi_{det}^i = \varphi_s^i - \phi_{ext}^i + \varphi_{add} \\
\phi_{det}^i = \left\{ 
\begin{array}{cl}
\varphi_m &\text{for} \ \varphi_{det}^i \leq \varphi_m \\
\varphi_{det}^i &\text{for} \ \varphi_m < \varphi_{det}^i \leq \TwoPi - \varphi_m\\
\varphi_m &\text{for} \ \TwoPi - \varphi_m < \varphi_{det}^i \leq \TwoPi + \varphi_m\\
\varphi_{det}^i - \TwoPi &\text{otherwise}\ .
\end{array}
 \right.
 \end{gathered}
\end{equation}
The phase margin $\varphi_{m}$ ensures that $\phi_{det}^i$ has a phase outside of the region $[-\varphi_m \ldots +\varphi_m]$ to avoid that the detection of the pulse is too close to the bus clock rising edge where the timing would be unreliable. It was chosen to be significantly larger than $\phi_{+}$.

The last parameters to be determined are the number of cycles each board has to wait before it can start output data on the bus. For this the propagation number of cycles $N_p^i$ have to be calculated:
\begin{equation}\label{eq:detection-time}
\begin{gathered}
N_{p}^i  = (\tau_p^i + t_g + t_d) \sslash T + \left\{ 
\begin{array}{cl}
N_0 &\text{for} \ \varphi_{det}^i \leq \TwoPi - \varphi_m\\
N_0 + 1 &\text{otherwise} \ .
\end{array}
 \right. 
 \end{gathered}
\end{equation}
Here the experimentally determined constant integer $N_0\,\epsilon\,\mathbb{Z}$ adds a few cycles to account for the cycles needed to start the output. The $+1$ accounts for the above mentioned case, that the detection clock was adjusted to detect the pulse one cycle later, to ensure reliable timing.
With the knowledge of all $N_{p}^i$ of the secondary boards the waiting number of cycles of the primary and secondary boards can be calculated:
\begin{equation}\label{eq:waiting-time}
\begin{aligned}
N_w^{prim} &= \text{max}_j(N_{p}^j)\\
N_w^i &= N_w^{prim} - N_{p}^i\ .
\end{aligned}
\end{equation}
The waiting number of cycles of the primary board is the largest of the $N_p^i$, i.e. max$_j(N_{p}^j)$, and each secondary board has to wait less until the last board does not need to wait.

The first demonstration of this scheme is presented in Sec. \ref{sub:auto-sync-data} and Fig. \ref{fig:auto-sync-data} shows the results. In Fig. \ref{fig:auto-sync-phase} in the next section the different phases are shown for the same data.

\subsection{Measured external clock phase}\label{sub:auto-sync-phase}

\begin{figure}[t]
\begin{center}
\includegraphics[width=\columnwidth]{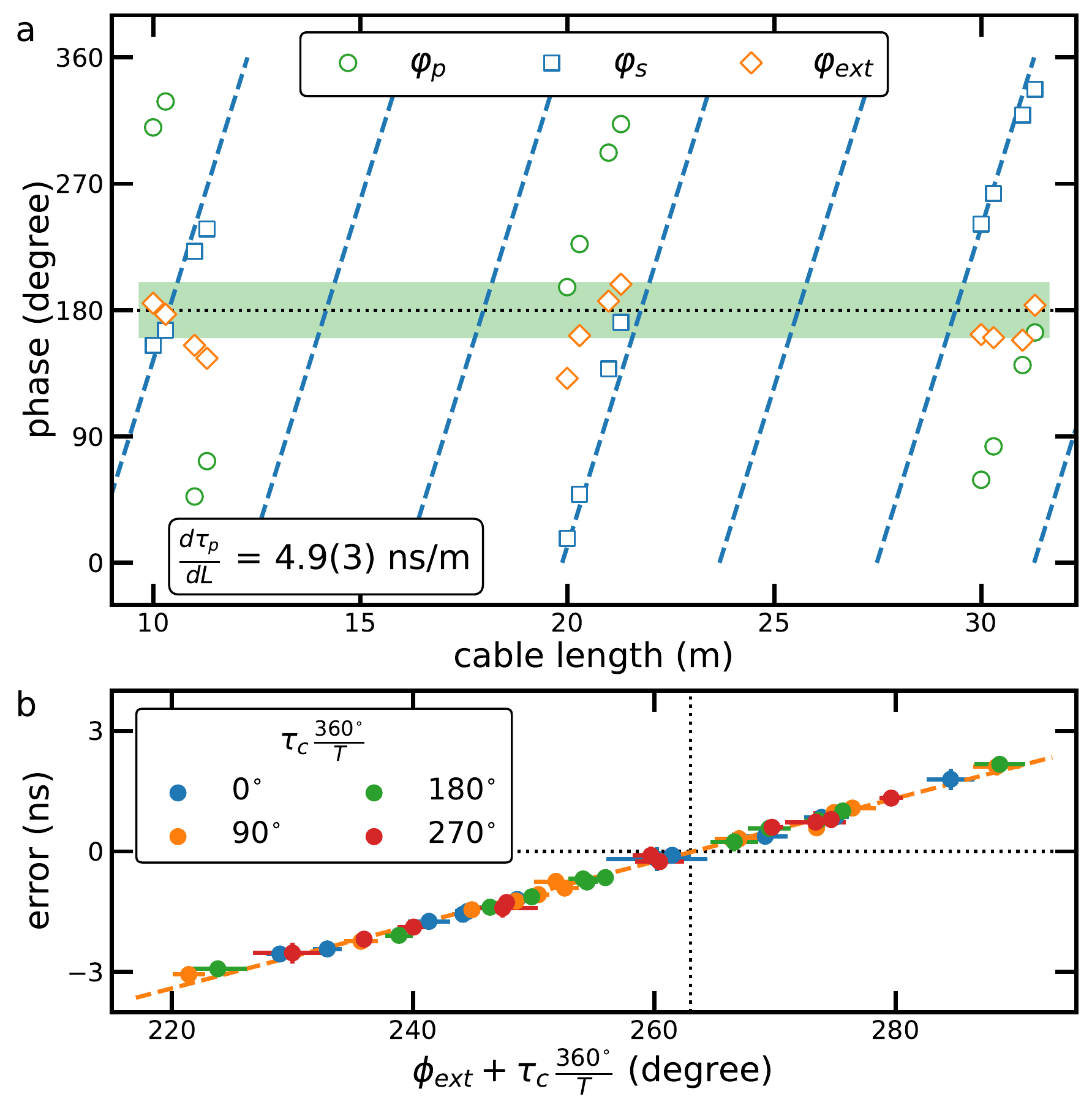}\\
\caption{a) Example phases for $\Delta \tau_c \frac{\TwoPi}{T} = 90^{\circ}$ corresponding to the data in Fig. \ref{fig:auto-sync-data}c: primary pulse phase ($\varphi_{p}$, green circles), secondary pulse phase ($\varphi_{s}$, blue squares) and external PLL phase ($\phi_{ext}$, orange diamonds). The dashed (blue) line is a linear fit modulo $\TwoPi$ through $\varphi_{s}$ and gives $\frac{d \tau_p}{d L}$ = 4.9(3)\,ns/m. For cable lengths 20\,m and 31.3\,m, the phase $\varphi_p$ is within $\pm 20^{\circ}$ of $\varphi_p^{crit} = 180^{\circ}$ (green shaded area) and the security phase is set nonzero $\xi(\varphi_p)$ = $\mp 30^{\circ}$. This causes that $\phi_{ext}$ is shifted away from the ideal value but ensures that the synchronization error, although slightly increased, does not jump arbitrarily by $\pm T$. b) Correlation between the error of the external clock phase ($\phi_{ext} + \Delta \tau_c \frac{\TwoPi}{T}$) and measured synchronization error plotted for all $\tau_c$ (different colors). The dashed (orange) line is a linear fit which gives a slope of 80(1)\,ps/degree and an offset of 263(4)$^{\circ}$.}
\label{fig:auto-sync-phase}
\end{center}
\end{figure}

Fig. \ref{fig:auto-sync-phase}a shows the phases $\varphi_p$ (green circles), $\varphi_{s}$ (blue squares) and $\phi_{ext}$ (orange diamonds) for the corresponding data presented in Fig. \ref{fig:auto-sync-data} in Sec. \ref{sub:auto-sync-data}.
The linear fit (modulo $\TwoPi$) of $\varphi_s$ vs. cable length (blue dashed line) gives a propagation delay per unit length of $\frac{d \tau_p}{d L}$ = 4.9(3)\,ns/m (averaged over leading and trailing edge of the pulse), which is the same as the one measured on the primary board (see Fig. \ref{fig:auto-sync-data}b). When $\Delta \tau_c = 0$, the offset of the linear fit gives the detection delay of the pulse $t_d$. For nonzero $\Delta \tau_c$ the offset is shifted accordingly.

In Fig. \ref{fig:auto-sync-phase}b we show the synchronization error as a function of the sum $\phi_{ext} + \Delta \tau_c \frac{\TwoPi}{T}$, i.e. how well the measured $\phi_{ext}$ compensates the externally applied clock delay $\Delta \tau_c$ (see Eq. \eqref{eq:clock-phase}). A linear fit (orange dashed line) gives a slope of 80(1)\,ps/degree which is slightly larger than the expected $\frac{20\,ns}{\TwoPi}$ = 53\,ps/degree and the offset of 263(4)$^{\circ}$ indicates that there is an additional unaccounted phase shift on the external clock. The used 20\,cm long clock cable would introduce a phase shift of only about 20$^{\circ}$ at the 50\,MHz external clock frequency used for this measurement. Additional phase shifts can come from input and clock buffers and propagation delays inside of the FPGA\footnote{We do not use the feedback option which cancels such phase shifts.}. The main contribution to the synchronization error can be attributed to the small difference of the measured pulse propagation delay per unit length $\frac{d \tau_p}{d L}$ between the primary board $\tau_p$ and the secondary board $\tau_s$. To compensate for this we have chosen to use the leading edge of the pulse on the primary board and the trailing edge on the secondary board. But with this choice the pulse width needs to be compensated (using $\phi_0$), which we do at the moment only under the assumption that it does not change for varying cable length. This assumption is not true due to the dispersion of the pulse. Nevertheless, even with the present scheme, the resulting synchronization error in Fig. \ref{fig:auto-sync-data}c is already very low.

\subsection{Fitting function for the synchronization error}\label{sub:trace-fit}

Here we present the fitting function used to fit the oscilloscope traces shown in the inset of Fig. \ref{fig:auto-sync-data}. For each trace the auto-synchronization was performed as described in Sec. \ref{sub:auto-sync-data}. After this, in order to measure the resulting synchronization error, another pulse is generated by the primary board and it waits the calculated waiting time $N_w^{prim}$ and generates a signal on an auxiliary I/O pin which is recorded by an oscilloscope (blue traces in inset of Fig. \ref{fig:auto-sync-data}). Each trace consists of 14 data points with a resolution of 2\,ns). After the secondary board detects the pulse, it immediately generates a signal on an auxiliary I/O pin which is used to trigger the oscilloscope and is recorded (orange traces) together with that of the primary board. The saved traces are fitted with a sigmoid function which is constructed from a piecewise defined linear slope $s(t,t_0,k,y_-,y_+)$ and is smoothed with a Gaussian kernel $g(t,\sigma)$:
\begin{equation}
\begin{aligned}
g(t,\sigma) &= \frac{1}{norm} e^{-\frac{t^2}{2 \sigma^2}}\\ 
\mu &= \frac{y_+ + y_-}{2},\ \ \nu = \frac{y_+ - y_-}{2\,k}\\
s(t,t_0,k,y_-,y_+) &= \left\{ 
\begin{array}{lll}
y_- &t - t_0 &\leq -\nu\\
\mu + (t-t_0)k \quad &|t - t_0| &< \nu\\
y_+ &t - t_0 &\ge \nu
\end{array}
\right. \\
f(t,t_0,k,\sigma,y_-,y_+) &= s(t,t_0,k,y_-,y_+) \star g(t,\sigma) \ .
\end{aligned}
\end{equation}
The symbol $\star$ means the discrete convolution with fixed steps in time and the Gaussian is normalized (norm) such that the sum over the discrete kernel entries is one. The function $f(t,t_0,k,\sigma,y_-,y_+)$ smoothly changes from the value $y_-$ for $t < t_0$ to the value $y_+$ for $t > t_0$. The slope $k$ and the width $\sigma$ of the Gaussian define how fast is the change between the extremes around the time $t_0$.

Each trace is fitted individually with $f(t,t_0,k,\sigma,y_-,y_+)$ with free parameters $t_0$, $k$, $y_-$ and $y_+$ and $\sigma = 2\,ns$ is kept fixed\footnote{When fitting $\sigma$, the correlation to the slope $k$ causes that for some traces the fit has problems to converge and attains big errors.}. The resulting synchronization error is the difference of the fitted $t_0^{sec}$ of the secondary board minus that one of the primary board $t_0^{prim}$.

\subsection{Measured detection signal}\label{sub:auto-sync-det}

\begin{figure}[t]
\begin{center}
\includegraphics[width=\columnwidth]{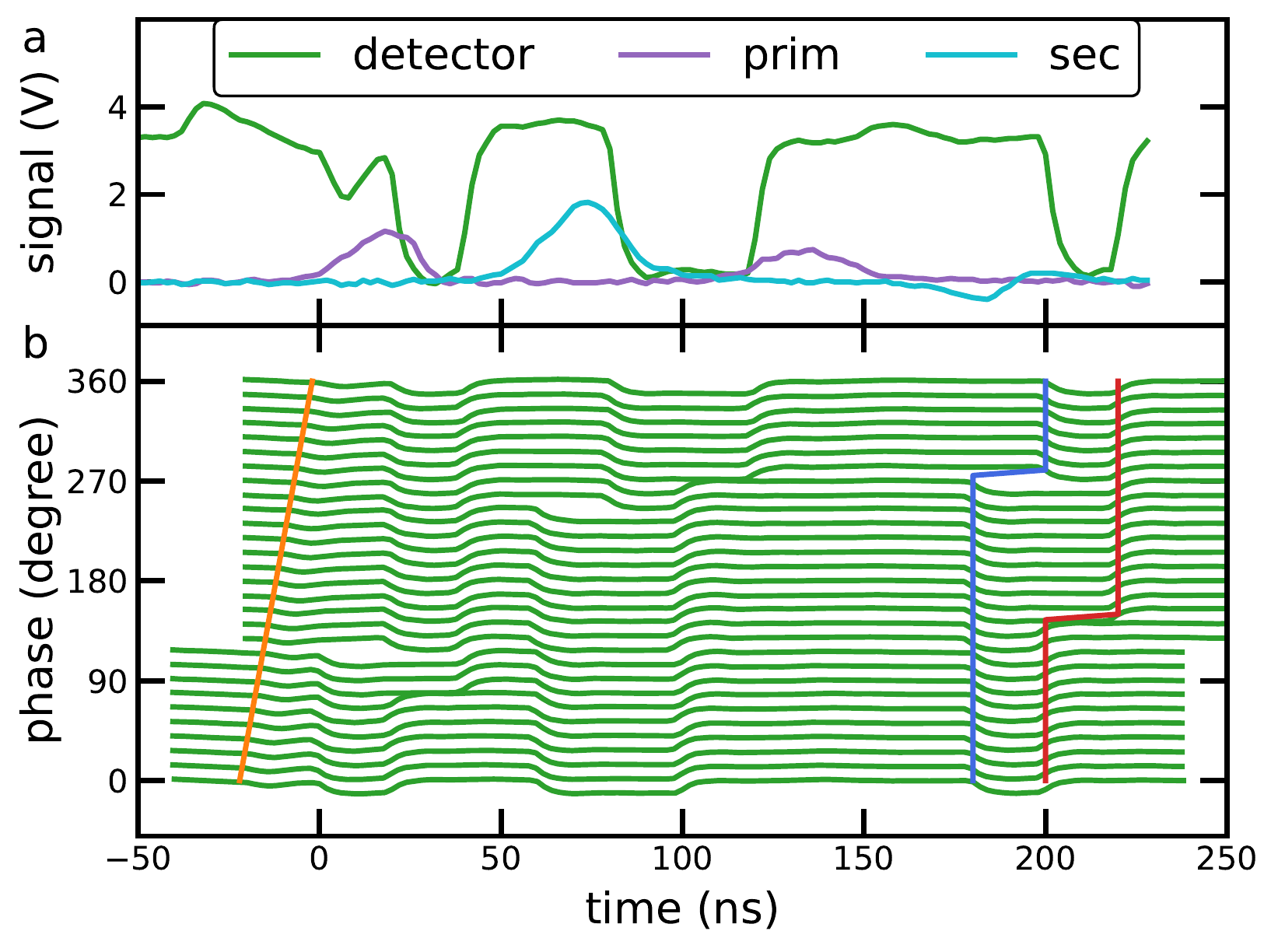}\\
\caption{Trigger coaxial and detector signals for 10\,m cable length measured on the primary board. a) Signals in the coaxial cable at the primary (violet) and secondary (cyan) boards for \TwoPi{} phase. The detector signal (green, active low) is generated by the primary FPGA when a pulse has been detected. 
b) Detector signals (green, offset by phase) for different phases. The pulse generation time (orange) is delayed linearly with phase and the detector signal shows jumps in the leading (blue) and trailing edge (red) of the reflected pulse.}
\label{fig:auto-sync-det}
\end{center}
\end{figure}

Here we show examples of trigger signals and the detection signal for varying detector phase used for the auto-synchronization described in Sec. \ref{sub:auto-sync}. The schematics of the pulse generation and detection electronics can be found in Ref. \onlinecite{github}. The present electronics was however designed for a first test and has not been optimized for efficiency and noise resilience. Additionally, it was designed for a test with two boards, where the 50 $\Omega$ termination is part of the generation and detection circuitry and a bipolar transistor, responsible for the reflection of the pulse, is inducing a high impedance in the coaxial cable instead of a short circuit as proposed.

Fig. \ref{fig:auto-sync-det}a shows the un-amplified signals in the trigger coaxial cable for primary (violet) and secondary (cyan) boards for \TwoPi{} phase and 10\,m cable length. The detector signal (green, active low) is generated by the primary board on an auxiliary I/O pin of the FPGA-SoC and indicates when the pulse has been detected after amplification and rectification by the FPGA-SoC. The first peak at 20\,ns is caused by noise on the supply when the pulse is created, the second at 80\,ns is the detection of the generated pulse, and the last peak at 200\,ns is the detection of the reflected pulse, which we are interested in. The delay of 3 cycles of these signals is caused by the required detector input synchronization stage (which is the same as a CDC) consisting of 2 flip-flops in series and one additional cycle to set or reset the output flip-flop. The small ripples on the signal is caused by the un-shielded and un-terminated clock signal cable used during this measurement.
Fig. \ref{fig:auto-sync-det}b shows the detector signal (green) at 10\,m cable length for different phases between the bus clock and the pulse. The time of the generation of the pulse is indicated by the orange line. The leading and trailing edges of the reflected pulse are indicated by the blue and red lines respectively. The jumps in these times are clearly visible and allow to measure the precise round-trip time with sub-cycle time resolution. See Fig. \ref{fig:auto-sync}b for comparison.

\section{Data rate fitting function}\label{sub:DMA-fit}

\begin{table}[tb]
\begin{minipage}{\linewidth}
\begin{tabular}{c|c|ccc|cc}
\hline
&&\multicolumn{3}{c|}{DMA}&\multicolumn{2}{c}{upload \& write}\\
&units&TX&RX&RX-FIFO&-10&-07S\\
\hline
max. $\Gamma$&MB/s&400\footnotemark[1]&300\footnotemark[1]&300\footnotemark[1]&118.7\footnotemark[2]&118.7\footnotemark[2]\\
\hline
$\tau$ &$\mu$s&0\footnotemark[3]&4.0(2)&0\footnotemark[3]&450(40)&350(50)\\
$N_{\Theta}$&1&4\footnotemark[3]&--&20(1)$\times10^3$&--&--\\  
$\Gamma_0$ &MB/s&600\footnotemark[3]&341(2)&600\footnotemark[3]&56.5(3)&47.2(4)\\
$\Gamma_1$ &MB/s&342.73(3)&--&340.50(5)&--&--\\
\hline
\end{tabular}
\caption{Fit results of the DMA and uploading data rates shown in figures \ref{fig:DMA} and \ref{fig:uploading} obtained with the model Eq. \eqref{eq:fit-model}. The DMA rates are given for the Cora-Z7-10 board but the rates of the Cora-Z7-07S board is within the error the same. The uploading rates include the writing to reserved DMA memory and is given for the Cora-Z7-10 and Cora-Z7-07S board (-10 and -07S in table headings respectively). The top row (max. $\Gamma$) gives the expected or theoretical maximum data rates in MB/s.} \label{tab:DMA-fit}

\footnotetext[1]{See Ref. \onlinecite{AXI-DMA} for the expected rates with the default DMA settings.}
\footnotetext[2]{See Ref. \onlinecite{GigE} for the maximum uploading rates. The measured rates include writing to reserved DMA memory.}
\footnotetext[3]{Fixed.}

\end{minipage}
\end{table}

Here we give the function used for modeling the measurements of the DMA transmission rates presented in Fig. \ref{fig:DMA}, Sec. \ref{sub:DMA}, and the data uploading rates presented in Fig. \ref{fig:uploading}, Sec. \ref{sub:uploading}. The fit results can be found in Tab. \ref{tab:DMA-fit}.

The model function gives the resulting rate $\Gamma(N)$ as a function of number of samples $N$ and includes a delay time (latency) $\tau$ and two data transmission rates where $\Gamma_0$ is active for $N \leq N_{\Theta}$ and $\Gamma_1$ active for $N > N_{\Theta}$:
\begin{equation}\label{eq:fit-model}
\begin{gathered}
\Gamma(N)  = \frac{1}{\frac{\tau}{N \beta} + \frac{\Theta(N_{\Theta}-N)}{\Gamma_0} + \frac{\Theta(N-N_{\Theta})}{\Gamma_1}} \\
\Theta(x) = \left\{ 
\begin{array}{cl}
0 &\text{for} \ x \leq 0\\
1 &\text{otherwise} \ .
\end{array}
 \right. 
 \end{gathered}
\end{equation}
The value $\beta =  12$ bytes per sample for this measurement. The delay takes into account that data cannot be transmitted immediately after the start signal has been given. The two rates are used to model that data transmission can run at different speeds, for example when FIFO buffers are involved.

For the measurement of the TX DMA rate an eventual delay cannot be detected and it was set to $\tau = 0$. The initial rate was set to the maximum possible $\Gamma_0 = \Gamma_{max}$ and the second rate $\Gamma_1$ is left as a fitting parameter. The threshold number of samples is set to fixed $N_{\Theta}$ = 4 since this is the smallest number of samples which can be transmitted. This is because we have chosen to use a 16 byte wide (128\,bits) data stream and $\beta = 12$ bytes, which have 48 bytes as the least common multiple, i.e. 4 samples.  Unused samples are marked by the driver with a ``no-operation'' (NOP) bit, such that non-multiple number of samples of 4 are no problem.
For the measurement of the RX DMA rate and the data uploading rate, the fitting parameters are the delay $\tau$ and the rate $\Gamma_0$. No second rate is needed.
For the measurement of the RX FIFO rate $\Gamma_1$ and $N_{\theta}$ are fitting parameters, the delay and initial rate is again set to 0 and $\Gamma_{max}$ respectively.

Note that the DMA rate measurements give not only the maximum possible bus output rate, but are as well an excellent tool to verify the efficiency of the driver. Any delays in time-critical parts, like the interrupt service routine or where the DMA buffers are updated, severely impact the DMA transmission rate. For example, output of text messages for debugging purposes cannot be done since the serial transmission of the text via USB to a host computer is too slow and would block the driver. 

\section{Start- and Stop trigger}\label{sub:start-stop}

\begin{figure}[t]
\begin{center}
\includegraphics[width=\columnwidth]{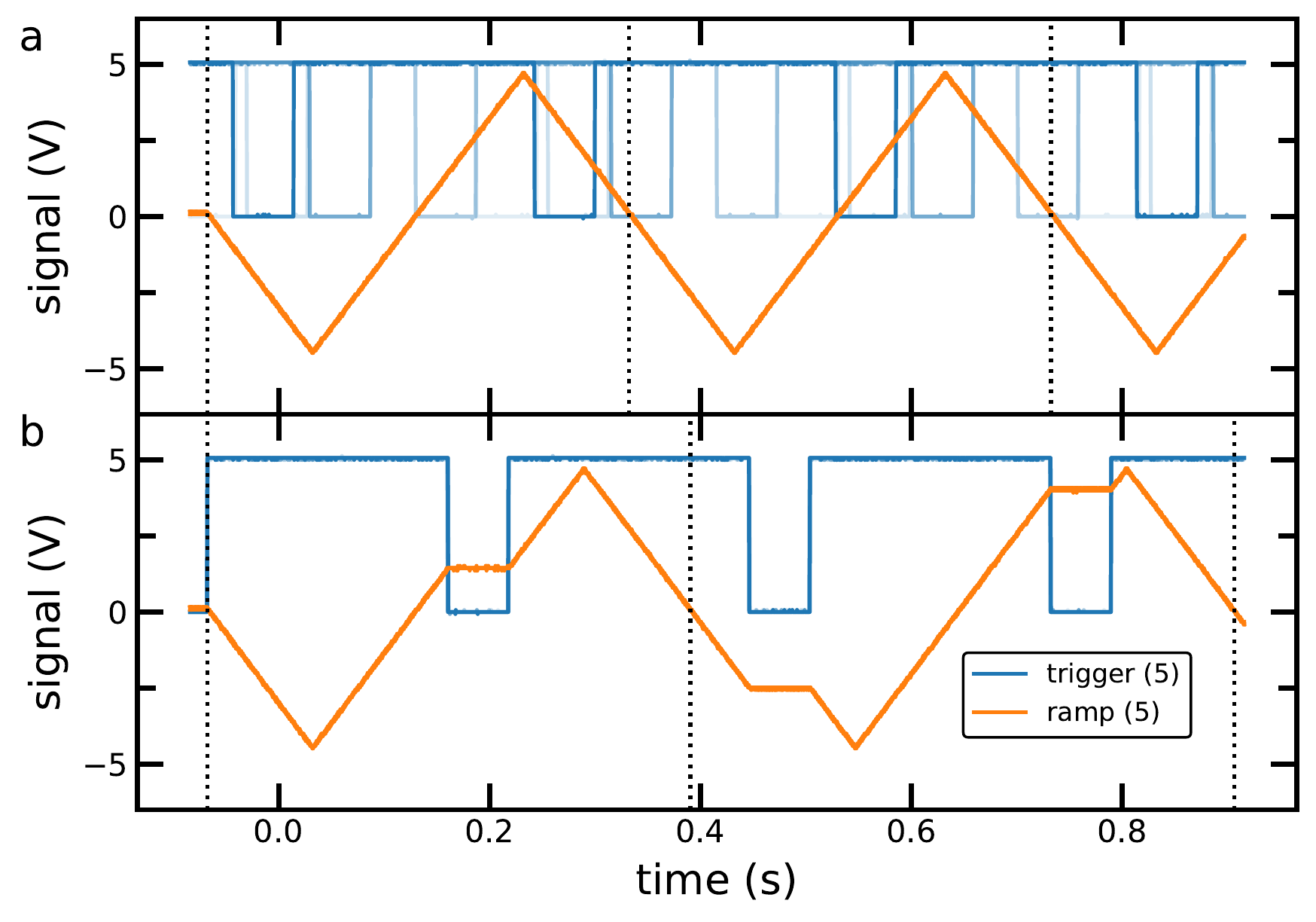}
\caption{Demonstration of a start and stop trigger option in cycling mode. a) Analog output triangular ramp (orange) running with 4\,$\mu$s 
per sample in cycling mode. The FPGA board is freely running without waiting for the trigger signal (blue). b) Same ramp but with start and stop trigger enabled. Both panels show the accumulated signal for 5 repetitions (number in brackets of labels). The vertical dotted lines indicate the beginning of each experimental cycle.}
\label{fig:start-stop}
\end{center}
\end{figure}

In Fig. \ref{fig:start-stop} we present a measurement of the start trigger and the cycling mode \footnote{In cycling mode the board repeats the experimental sequence for a programmed number of times or infinitely until a stop command is sent.} of the board. 
In addition, we implemented for demonstration the possibility to interrupt the execution of the sequence when the start trigger signal is reset after the board has been started. This might be useful to manually check the state of the experiment, or to wait for some external event, like waiting until the atom number reaches a certain value. The experimental sequence consists of an analog output performing a triangular ramp (orange) which is executed repeatedly in cycling mode. The dotted lines indicate the beginning of each cycle. A waveform generator provides the trigger signal (blue). See Fig. \ref{fig:start-stop}a for the unperturbed experiment: without the start-stop trigger activated, there is no relation between the trigger and the ramp, which we show for 5 realisations of the experiment. In Fig. \ref{fig:start-stop}b we show the result when the start-stop trigger is activated which is starting the execution of the ramp and then interrupting it as long as the trigger signal is low. We have again repeated this measurement 5 times and now all repetitions overlap. 

\section{Resource utilization}\label{sub:resources}

Tab. \ref{tab:resources} gives a summary of the used resources of the PL part and shows that we do not use all of the available resources although the FPGA is relatively small. This allows to implement further improvements or customization in case it is needed.

\begin{table}[tb]
\begin{minipage}{\columnwidth}
\begin{tabular}{c|l|cccccc}
device&&FF &LUT &BRAM &MMCM&PLL&DSP\\
\hline
\multirow{3}{*}{\rotatebox[origin=c]{90}{Z7-10}}&available&35200 &17600&60&2&2&80\\
\cline{2-8}
&used&13275 &9824 &38 &2&0&0\\
&percent &38 &56 &63 &100&0&0\\
\hline\hline
\multirow{3}{*}{\rotatebox[origin=c]{90}{Z7-07S}}&available&28800 &14400&50&2&2&66\\
\cline{2-8}
&used&13274 &9825 &38 &2&0&0\\
&percent &46 &68 &76 &100&0&0\\
\hline
\end{tabular}
\caption{Used resources for Cora Z7-10 (Zynq-7010) and Cora Z7-07S (Zynq-7007S) FPGA-SoC boards: Flip-flops (FF) store single-bit data, lookup tables (LUT) are used to represent logic operations, block-RAM (BRAM) is a much larger (36\,kbit) collection of flip-flops, and mixed-mode manager (MMCM), are phase-locked loops (PLL) but allow dynamic phase shifting. We use neither classical PLLs nor digital signal processing (DSP) cells.}
\label{tab:resources}
\end{minipage}
\end{table}

\nocite{*}
\bibliography{aipsamp}

\providecommand{\noopsort}[1]{}\providecommand{\singleletter}[1]{#1}%
\begin{thebibliography}{65}%
\makeatletter
\providecommand \@ifxundefined [1]{%
 \@ifx{#1\undefined}
}%
\providecommand \@ifnum [1]{%
 \ifnum #1\expandafter \@firstoftwo
 \else \expandafter \@secondoftwo
 \fi
}%
\providecommand \@ifx [1]{%
 \ifx #1\expandafter \@firstoftwo
 \else \expandafter \@secondoftwo
 \fi
}%
\providecommand \natexlab [1]{#1}%
\providecommand \enquote  [1]{``#1''}%
\providecommand \bibnamefont  [1]{#1}%
\providecommand \bibfnamefont [1]{#1}%
\providecommand \citenamefont [1]{#1}%
\providecommand \href@noop [0]{\@secondoftwo}%
\providecommand \href [0]{\begingroup \@sanitize@url \@href}%
\providecommand \@href[1]{\@@startlink{#1}\@@href}%
\providecommand \@@href[1]{\endgroup#1\@@endlink}%
\providecommand \@sanitize@url [0]{\catcode `\\12\catcode `\$12\catcode
  `\&12\catcode `\#12\catcode `\^12\catcode `\_12\catcode `\%12\relax}%
\providecommand \@@startlink[1]{}%
\providecommand \@@endlink[0]{}%
\providecommand \url  [0]{\begingroup\@sanitize@url \@url }%
\providecommand \@url [1]{\endgroup\@href {#1}{\urlprefix }}%
\providecommand \urlprefix  [0]{URL }%
\providecommand \Eprint [0]{\href }%
\providecommand \doibase [0]{http://dx.doi.org/}%
\providecommand \selectlanguage [0]{\@gobble}%
\providecommand \bibinfo  [0]{\@secondoftwo}%
\providecommand \bibfield  [0]{\@secondoftwo}%
\providecommand \translation [1]{[#1]}%
\providecommand \BibitemOpen [0]{}%
\providecommand \bibitemStop [0]{}%
\providecommand \bibitemNoStop [0]{.\EOS\space}%
\providecommand \EOS [0]{\spacefactor3000\relax}%
\providecommand \BibitemShut  [1]{\csname bibitem#1\endcsname}%
\let\auto@bib@innerbib\@empty
\bibitem [{NI-()}]{NI-FPGA}%
  \BibitemOpen
  \href
  {https://www.ni.com/en-us/shop/hardware/products/digital-reconfigurable-io-device.html}
  {}\bibinfo {note} {National Instruments Digital Reconfigurable I/O Device,
  \url{https://www.ni.com/en-us/shop/hardware/products/digital-reconfigurable-io-device.html}}\BibitemShut
  {NoStop}%
\bibitem [{Art()}]{Artiq2007}%
  \BibitemOpen
  \href {https://m-labs.hk/experiment-control/artiq/} {}\bibinfo {note} {ARTIQ,
  open-source experimental control system,
  \url{https://m-labs.hk/experiment-control/artiq/}}\BibitemShut {NoStop}%
\bibitem [{\citenamefont {Keshet}\ and\ \citenamefont
  {Ketterle}(2013)}]{Ketterle2013}%
  \BibitemOpen
  \bibfield  {author} {\bibinfo {author} {\bibfnamefont {A.}~\bibnamefont
  {Keshet}}\ and\ \bibinfo {author} {\bibfnamefont {W.}~\bibnamefont
  {Ketterle}},\ }\href {\doibase 10.1063/1.4773536} {\bibfield  {journal}
  {\bibinfo  {journal} {Rev. Sci. Instrum.}\ }\textbf {\bibinfo {volume}
  {84}},\ \bibinfo {pages} {015105} (\bibinfo {year} {2013})}\BibitemShut
  {NoStop}%
\bibitem [{\citenamefont {Ramola}(2015)}]{Meschede2015}%
  \BibitemOpen
  \bibfield  {author} {\bibinfo {author} {\bibfnamefont {G.}~\bibnamefont
  {Ramola}},\ }\emph {\bibinfo {title} {A versatile digital frequency
  synthesizer for state-dependent transport of trapped neutral atoms}},\ \href
  {http://quantum-technologies.iap.uni-bonn.de/en/diplom-theses.html?task=download&file=293&token=4a147f8b64fbf27b309b217f0d3729eb}
  {\bibinfo {type} {Master thesis}},\ \bibinfo  {school} {Rheinischen
  Friedrich-Wilhelms-Universit{\"a}t Bonn} (\bibinfo {year} {2015})\BibitemShut
  {NoStop}%
\bibitem [{\citenamefont {Pruttivarasin}\ and\ \citenamefont
  {Katori}(2015)}]{Katori2015}%
  \BibitemOpen
  \bibfield  {author} {\bibinfo {author} {\bibfnamefont {T.}~\bibnamefont
  {Pruttivarasin}}\ and\ \bibinfo {author} {\bibfnamefont {H.}~\bibnamefont
  {Katori}},\ }\href {\doibase 10.1063/1.4935476} {\bibfield  {journal}
  {\bibinfo  {journal} {Rev. Sci. Instrum.}\ }\textbf {\bibinfo {volume}
  {86}},\ \bibinfo {pages} {115106} (\bibinfo {year} {2015})}\BibitemShut
  {NoStop}%
\bibitem [{\citenamefont {Du}\ \emph {et~al.}(2017)\citenamefont {Du},
  \citenamefont {Li}, \citenamefont {Ge}, \citenamefont {Lu}, \citenamefont
  {Deng},\ and\ \citenamefont {Lu}}]{Lu2017}%
  \BibitemOpen
  \bibfield  {author} {\bibinfo {author} {\bibfnamefont {Y.}~\bibnamefont
  {Du}}, \bibinfo {author} {\bibfnamefont {W.}~\bibnamefont {Li}}, \bibinfo
  {author} {\bibfnamefont {Y.}~\bibnamefont {Ge}}, \bibinfo {author}
  {\bibfnamefont {H.}~\bibnamefont {Lu}}, \bibinfo {author} {\bibfnamefont
  {K.}~\bibnamefont {Deng}}, \ and\ \bibinfo {author} {\bibfnamefont
  {Z.}~\bibnamefont {Lu}},\ }\href {\doibase 10.1063/1.5001489} {\bibfield
  {journal} {\bibinfo  {journal} {Rev. Sci. Instrum.}\ }\textbf {\bibinfo
  {volume} {88}},\ \bibinfo {pages} {096103} (\bibinfo {year}
  {2017})}\BibitemShut {NoStop}%
\bibitem [{\citenamefont {Donnellan}\ \emph {et~al.}(2019)\citenamefont
  {Donnellan}, \citenamefont {Hill}, \citenamefont {Bowden},\ and\
  \citenamefont {Hobson}}]{Hobson2019}%
  \BibitemOpen
  \bibfield  {author} {\bibinfo {author} {\bibfnamefont {S.}~\bibnamefont
  {Donnellan}}, \bibinfo {author} {\bibfnamefont {I.~R.}\ \bibnamefont {Hill}},
  \bibinfo {author} {\bibfnamefont {W.}~\bibnamefont {Bowden}}, \ and\ \bibinfo
  {author} {\bibfnamefont {R.}~\bibnamefont {Hobson}},\ }\href {\doibase
  10.1063/1.5051124} {\bibfield  {journal} {\bibinfo  {journal} {Rev. Sci.
  Instrum.}\ }\textbf {\bibinfo {volume} {90}},\ \bibinfo {pages} {043101}
  (\bibinfo {year} {2019})}\BibitemShut {NoStop}%
\bibitem [{\citenamefont {Mattingly}\ and\ \citenamefont
  {Skiff}(2018)}]{LIF2018}%
  \BibitemOpen
  \bibfield  {author} {\bibinfo {author} {\bibfnamefont {S.~W.}\ \bibnamefont
  {Mattingly}}\ and\ \bibinfo {author} {\bibfnamefont {F.}~\bibnamefont
  {Skiff}},\ }\href {\doibase 10.1063/1.4995971} {\bibfield  {journal}
  {\bibinfo  {journal} {Rev. Sci. Instrum.}\ }\textbf {\bibinfo {volume}
  {89}},\ \bibinfo {pages} {043508} (\bibinfo {year} {2018})}\BibitemShut
  {NoStop}%
\bibitem [{\citenamefont {Shu}\ \emph {et~al.}(2018)\citenamefont {Shu},
  \citenamefont {Wang}, \citenamefont {Liu}, \citenamefont {Meiwen},
  \citenamefont {Zhang}, \citenamefont {Jiarong},\ and\ \citenamefont
  {Ji}}]{Tokamak2018}%
  \BibitemOpen
  \bibfield  {author} {\bibinfo {author} {\bibfnamefont {S.}~\bibnamefont
  {Shu}}, \bibinfo {author} {\bibfnamefont {L.}~\bibnamefont {Wang}}, \bibinfo
  {author} {\bibfnamefont {D.}~\bibnamefont {Liu}}, \bibinfo {author}
  {\bibfnamefont {C.}~\bibnamefont {Meiwen}}, \bibinfo {author} {\bibfnamefont
  {Y.}~\bibnamefont {Zhang}}, \bibinfo {author} {\bibfnamefont
  {L.}~\bibnamefont {Jiarong}}, \ and\ \bibinfo {author} {\bibfnamefont
  {F.}~\bibnamefont {Ji}},\ }\href {\doibase 10.1063/1.5035364} {\bibfield
  {journal} {\bibinfo  {journal} {Rev. Sci. Instrum.}\ }\textbf {\bibinfo
  {volume} {89}} (\bibinfo {year} {2018}),\ 10.1063/1.5035364}\BibitemShut
  {NoStop}%
\bibitem [{\citenamefont {Perego}\ \emph {et~al.}(2018)\citenamefont {Perego},
  \citenamefont {Pomponio}, \citenamefont {Detti}, \citenamefont {Duca},
  \citenamefont {Sias},\ and\ \citenamefont {Calosso}}]{Sias2018}%
  \BibitemOpen
  \bibfield  {author} {\bibinfo {author} {\bibfnamefont {E.}~\bibnamefont
  {Perego}}, \bibinfo {author} {\bibfnamefont {M.}~\bibnamefont {Pomponio}},
  \bibinfo {author} {\bibfnamefont {A.}~\bibnamefont {Detti}}, \bibinfo
  {author} {\bibfnamefont {L.}~\bibnamefont {Duca}}, \bibinfo {author}
  {\bibfnamefont {C.}~\bibnamefont {Sias}}, \ and\ \bibinfo {author}
  {\bibfnamefont {C.~E.}\ \bibnamefont {Calosso}},\ }\href {\doibase
  10.1063/1.5049120} {\bibfield  {journal} {\bibinfo  {journal} {Rev. Sci.
  Instrum.}\ }\textbf {\bibinfo {volume} {89}},\ \bibinfo {pages} {113116}
  (\bibinfo {year} {2018})}\BibitemShut {NoStop}%
\bibitem [{\citenamefont {Yu}\ \emph {et~al.}(2018)\citenamefont {Yu},
  \citenamefont {Fajeau}, \citenamefont {Liu}, \citenamefont {Jones},\ and\
  \citenamefont {Madison}}]{Madison2018}%
  \BibitemOpen
  \bibfield  {author} {\bibinfo {author} {\bibfnamefont {S.~J.}\ \bibnamefont
  {Yu}}, \bibinfo {author} {\bibfnamefont {E.}~\bibnamefont {Fajeau}}, \bibinfo
  {author} {\bibfnamefont {L.~Q.}\ \bibnamefont {Liu}}, \bibinfo {author}
  {\bibfnamefont {D.~J.}\ \bibnamefont {Jones}}, \ and\ \bibinfo {author}
  {\bibfnamefont {K.~W.}\ \bibnamefont {Madison}},\ }\href {\doibase
  10.1063/1.5001312} {\bibfield  {journal} {\bibinfo  {journal} {Rev. Sci.
  Instrum.}\ }\textbf {\bibinfo {volume} {89}},\ \bibinfo {pages} {025107}
  (\bibinfo {year} {2018})}\BibitemShut {NoStop}%
\bibitem [{\citenamefont {Ristè}\ \emph {et~al.}(2013)\citenamefont {Ristè},
  \citenamefont {Dukalski}, \citenamefont {Watson}, \citenamefont {de~Lange},
  \citenamefont {Tiggelman}, \citenamefont {Blanter}, \citenamefont {Lehnert},
  \citenamefont {Schouten},\ and\ \citenamefont
  {DiCarlo}}]{QuantumFeedback2013}%
  \BibitemOpen
  \bibfield  {author} {\bibinfo {author} {\bibfnamefont {D.}~\bibnamefont
  {Ristè}}, \bibinfo {author} {\bibfnamefont {M.}~\bibnamefont {Dukalski}},
  \bibinfo {author} {\bibfnamefont {C.~A.}\ \bibnamefont {Watson}}, \bibinfo
  {author} {\bibfnamefont {G.}~\bibnamefont {de~Lange}}, \bibinfo {author}
  {\bibfnamefont {M.~J.}\ \bibnamefont {Tiggelman}}, \bibinfo {author}
  {\bibfnamefont {Y.~M.}\ \bibnamefont {Blanter}}, \bibinfo {author}
  {\bibfnamefont {K.~W.}\ \bibnamefont {Lehnert}}, \bibinfo {author}
  {\bibfnamefont {R.~N.}\ \bibnamefont {Schouten}}, \ and\ \bibinfo {author}
  {\bibfnamefont {L.}~\bibnamefont {DiCarlo}},\ }\href {\doibase
  10.1038/nature12513} {\bibfield  {journal} {\bibinfo  {journal} {Nature}\
  }\textbf {\bibinfo {volume} {502}},\ \bibinfo {pages} {350} (\bibinfo {year}
  {2013})}\BibitemShut {NoStop}%
\bibitem [{\citenamefont {Lamb}\ \emph {et~al.}(2016)\citenamefont {Lamb},
  \citenamefont {Colless}, \citenamefont {Hornibrook}, \citenamefont {Pauka},
  \citenamefont {Waddy}, \citenamefont {Frechtling},\ and\ \citenamefont
  {Reilly}}]{cryogenic2016}%
  \BibitemOpen
  \bibfield  {author} {\bibinfo {author} {\bibfnamefont {I.}~\bibnamefont
  {Lamb}}, \bibinfo {author} {\bibfnamefont {J.}~\bibnamefont {Colless}},
  \bibinfo {author} {\bibfnamefont {J.}~\bibnamefont {Hornibrook}}, \bibinfo
  {author} {\bibfnamefont {S.}~\bibnamefont {Pauka}}, \bibinfo {author}
  {\bibfnamefont {S.}~\bibnamefont {Waddy}}, \bibinfo {author} {\bibfnamefont
  {M.}~\bibnamefont {Frechtling}}, \ and\ \bibinfo {author} {\bibfnamefont
  {D.}~\bibnamefont {Reilly}},\ }\href {\doibase 10.1063/1.4939094} {\bibfield
  {journal} {\bibinfo  {journal} {Rev. Sci. Instrum.}\ }\textbf {\bibinfo
  {volume} {87}},\ \bibinfo {pages} {014701} (\bibinfo {year}
  {2016})}\BibitemShut {NoStop}%
\bibitem [{\citenamefont {Homulle}\ \emph {et~al.}(2017)\citenamefont
  {Homulle}, \citenamefont {Visser}, \citenamefont {Patra}, \citenamefont
  {Ferrari}, \citenamefont {Prati}, \citenamefont {Sebastiano},\ and\
  \citenamefont {Charbon}}]{cryogenic2017}%
  \BibitemOpen
  \bibfield  {author} {\bibinfo {author} {\bibfnamefont {H.}~\bibnamefont
  {Homulle}}, \bibinfo {author} {\bibfnamefont {S.}~\bibnamefont {Visser}},
  \bibinfo {author} {\bibfnamefont {B.}~\bibnamefont {Patra}}, \bibinfo
  {author} {\bibfnamefont {G.}~\bibnamefont {Ferrari}}, \bibinfo {author}
  {\bibfnamefont {E.}~\bibnamefont {Prati}}, \bibinfo {author} {\bibfnamefont
  {F.}~\bibnamefont {Sebastiano}}, \ and\ \bibinfo {author} {\bibfnamefont
  {E.}~\bibnamefont {Charbon}},\ }\href {\doibase 10.1063/1.4979611} {\bibfield
   {journal} {\bibinfo  {journal} {Rev. Sci. Instrum.}\ }\textbf {\bibinfo
  {volume} {88}},\ \bibinfo {pages} {045103} (\bibinfo {year}
  {2017})}\BibitemShut {NoStop}%
\bibitem [{\citenamefont {Qin}\ \emph {et~al.}(2020)\citenamefont {Qin},
  \citenamefont {Zhang}, \citenamefont {Wang}, \citenamefont {Zhao},
  \citenamefont {Tong}, \citenamefont {Rong},\ and\ \citenamefont
  {Du}}]{SpinControl2020}%
  \BibitemOpen
  \bibfield  {author} {\bibinfo {author} {\bibfnamefont {X.}~\bibnamefont
  {Qin}}, \bibinfo {author} {\bibfnamefont {W.}~\bibnamefont {Zhang}}, \bibinfo
  {author} {\bibfnamefont {L.}~\bibnamefont {Wang}}, \bibinfo {author}
  {\bibfnamefont {Y.}~\bibnamefont {Zhao}}, \bibinfo {author} {\bibfnamefont
  {Y.}~\bibnamefont {Tong}}, \bibinfo {author} {\bibfnamefont {X.}~\bibnamefont
  {Rong}}, \ and\ \bibinfo {author} {\bibfnamefont {J.}~\bibnamefont {Du}},\
  }\href {\doibase 10.1109/TIM.2019.2910921} {\bibfield  {journal} {\bibinfo
  {journal} {IEEE Trans. Instrum. Meas.}\ }\textbf {\bibinfo {volume} {69}},\
  \bibinfo {pages} {1127} (\bibinfo {year} {2020})}\BibitemShut {NoStop}%
\bibitem [{\citenamefont {Xu}\ \emph {et~al.}(2021)\citenamefont {Xu},
  \citenamefont {Huang}, \citenamefont {Balewski}, \citenamefont {Naik},
  \citenamefont {Morvan}, \citenamefont {Mitchell}, \citenamefont {Nowrouzi},
  \citenamefont {Santiago},\ and\ \citenamefont {Siddiqi}}]{QubiC2021}%
  \BibitemOpen
  \bibfield  {author} {\bibinfo {author} {\bibfnamefont {Y.}~\bibnamefont
  {Xu}}, \bibinfo {author} {\bibfnamefont {G.}~\bibnamefont {Huang}}, \bibinfo
  {author} {\bibfnamefont {J.}~\bibnamefont {Balewski}}, \bibinfo {author}
  {\bibfnamefont {R.}~\bibnamefont {Naik}}, \bibinfo {author} {\bibfnamefont
  {A.}~\bibnamefont {Morvan}}, \bibinfo {author} {\bibfnamefont
  {B.}~\bibnamefont {Mitchell}}, \bibinfo {author} {\bibfnamefont
  {K.}~\bibnamefont {Nowrouzi}}, \bibinfo {author} {\bibfnamefont {D.~I.}\
  \bibnamefont {Santiago}}, \ and\ \bibinfo {author} {\bibfnamefont
  {I.}~\bibnamefont {Siddiqi}},\ }\href {https://arxiv.org/abs/2101.00071}
  {\bibfield  {journal} {\bibinfo  {journal} {arXiv}\ } (\bibinfo {year}
  {2021})},\ \bibinfo {note} {arxiv.org/abs/2101.00071}\BibitemShut {NoStop}%
\bibitem [{\citenamefont {Habinc}(2002)}]{FPGA_in_space}%
  \BibitemOpen
  \bibfield  {author} {\bibinfo {author} {\bibfnamefont {S.}~\bibnamefont
  {Habinc}},\ }\href
  {http://microelectronics.esa.int/techno/fpga_002_01-0-4.pdf} {\emph {\bibinfo
  {title} {Suitability of reprogrammable FPGAs in space applications}}}\
  (\bibinfo  {publisher} {Gaisler Research},\ \bibinfo {year} {2002})\ \bibinfo
  {note}
  {\url{http://microelectronics.esa.int/techno/fpga_002_01-0-4.pdf}}\BibitemShut
  {NoStop}%
\bibitem [{\citenamefont {Bertoldi}\ \emph {et~al.}(2020)\citenamefont
  {Bertoldi}, \citenamefont {Feng}, \citenamefont {Eneriz}, \citenamefont
  {Carey}, \citenamefont {Naik}, \citenamefont {Junca}, \citenamefont {Zou},
  \citenamefont {Sabulsky}, \citenamefont {Canuel}, \citenamefont {Bouyer},\
  and\ \citenamefont {Prevedelli}}]{Prevedelli2020}%
  \BibitemOpen
  \bibfield  {author} {\bibinfo {author} {\bibfnamefont {A.}~\bibnamefont
  {Bertoldi}}, \bibinfo {author} {\bibfnamefont {C.-H.}\ \bibnamefont {Feng}},
  \bibinfo {author} {\bibfnamefont {H.}~\bibnamefont {Eneriz}}, \bibinfo
  {author} {\bibfnamefont {M.}~\bibnamefont {Carey}}, \bibinfo {author}
  {\bibfnamefont {D.~S.}\ \bibnamefont {Naik}}, \bibinfo {author}
  {\bibfnamefont {Z.}~\bibnamefont {Junca}}, \bibinfo {author} {\bibfnamefont
  {X.}~\bibnamefont {Zou}}, \bibinfo {author} {\bibfnamefont {D.~O.}\
  \bibnamefont {Sabulsky}}, \bibinfo {author} {\bibfnamefont {B.}~\bibnamefont
  {Canuel}}, \bibinfo {author} {\bibfnamefont {P.}~\bibnamefont {Bouyer}}, \
  and\ \bibinfo {author} {\bibfnamefont {M.}~\bibnamefont {Prevedelli}},\
  }\href {\doibase 10.1063/1.5129595} {\bibfield  {journal} {\bibinfo
  {journal} {Rev. Sci. Instrum.}\ }\textbf {\bibinfo {volume} {91}},\ \bibinfo
  {pages} {033203} (\bibinfo {year} {2020})}\BibitemShut {NoStop}%
\bibitem [{Whi()}]{WhiteRabbit}%
  \BibitemOpen
  \href {https://white-rabbit.web.cern.ch/} {}\bibinfo {note} {CERN, The White
  Rabbit Project, \url{https://white-rabbit.web.cern.ch/}}\BibitemShut
  {NoStop}%
\bibitem [{Cor()}]{Cora}%
  \BibitemOpen
  \href
  {https://reference.digilentinc.com/reference/programmable-logic/cora-z7/start}
  {}\bibinfo {note} {Cora-Z7-10 and Cora-Z7-07S development boards from
  Digilent Inc.,
  \url{https://reference.digilentinc.com/reference/programmable-logic/cora-z7/start}}\BibitemShut
  {NoStop}%
\bibitem [{Ter()}]{TerasicDE10Nano}%
  \BibitemOpen
  \href
  {https://www.terasic.com.tw/cgi-bin/page/archive.pl?Language=English&CategoryNo=167&No=1046}
  {}\bibinfo {note} {DE10-Nano Kit from Terasic Inc.,
  \url{https://www.terasic.com.tw/cgi-bin/page/archive.pl?Language=English&CategoryNo=167&No=1046}}\BibitemShut
  {NoStop}%
\bibitem [{git()}]{github}%
  \BibitemOpen
  \href {https://github.com/INO-quantum/FPGA-SoC-experiment-control} {}\bibinfo
  {note} {The source code, electronic schemes and Gerber files, the
  instructions for installation of the software and the compilation of the
  sources can be found at
  \url{https://github.com/INO-quantum/FPGA-SoC-experiment-control}.}\BibitemShut
  {Stop}%
\bibitem [{\citenamefont {Cummings}(2008)}]{CummingsCDC2008}%
  \BibitemOpen
  \bibfield  {author} {\bibinfo {author} {\bibfnamefont {C.~E.}\ \bibnamefont
  {Cummings}},\ }\href
  {https://www.sunburst-design.com/papers/CummingsSNUG2008Boston_CDC.pdf}
  {\enquote {\bibinfo {title} {Clock {D}omain {C}rossings ({CDC}) {D}esign \&
  {V}erification {T}echniques {U}sing {S}ystem {V}erilog},}\ } (\bibinfo {year}
  {2008}),\ \bibinfo {note} {{SNUG} 2008, Boston.
  \url{https://www.sunburst-design.com/papers/CummingsSNUG2008Boston_CDC.pdf}}\BibitemShut
  {NoStop}%
\bibitem [{AMB()}]{AMBA-AXI}%
  \BibitemOpen
  \href {https://developer.arm.com/documentation/ihi0022/latest/} {}\bibinfo
  {note} {Second release of AMBA AXI and ACE Protocol Specification, Issue E,
  22 February 2013. The Advanced eXtensible Interface (AXI) protocol is a part
  of ARM Advanced Microcontroller Bus (AMBA) structure.
  \url{https://developer.arm.com/documentation/ihi0022/latest/}}\BibitemShut
  {NoStop}%
\bibitem [{AXI()}]{AXI-DMA}%
  \BibitemOpen
  \href
  {https://www.xilinx.com/support/documentation/ip_documentation/axi_dma/v7_1/pg021_axi_dma.pdf}
  {}\bibinfo {note} {Xilinx AXI DMA v7.1 LogicCORE IP product guide, PG021,
  June 14 2019,
  \url{https://www.xilinx.com/support/documentation/ip_documentation/axi_dma/v7_1/pg021_axi_dma.pdf}}\BibitemShut
  {NoStop}%
\bibitem [{\citenamefont {Cummings}(2002)}]{CummingsFIFO2002a}%
  \BibitemOpen
  \bibfield  {author} {\bibinfo {author} {\bibfnamefont {C.~E.}\ \bibnamefont
  {Cummings}},\ }\href
  {https://www.sunburst-design.com/papers/CummingsSNUG2002SJ_FIFO1.pdf}
  {\enquote {\bibinfo {title} {Simulation and {S}ynthesis {T}echniques for
  {A}synchronous {FIFO} design},}\ } (\bibinfo {year} {2002}),\ \bibinfo {note}
  {{SNUG} 2002, San Jose.
  \url{https://www.sunburst-design.com/papers/CummingsSNUG2002SJ_FIFO1.pdf}}\BibitemShut
  {NoStop}%
\bibitem [{\citenamefont {Cummings}\ and\ \citenamefont
  {Alfke}(2002)}]{CummingsFIFO2002b}%
  \BibitemOpen
  \bibfield  {author} {\bibinfo {author} {\bibfnamefont {C.~E.}\ \bibnamefont
  {Cummings}}\ and\ \bibinfo {author} {\bibfnamefont {P.}~\bibnamefont
  {Alfke}},\ }\href
  {https://www.sunburst-design.com/papers/CummingsSNUG2002SJ_FIFO1.pdf}
  {\enquote {\bibinfo {title} {Simulation and {S}ynthesis techniques for
  {A}synchronous {FIFO} {D}esign with {A}synchronous {P}ointer
  {C}omparisons},}\ } (\bibinfo {year} {2002}),\ \bibinfo {note} {{SNUG} 2002,
  San Jose.
  \url{https://www.sunburst-design.com/papers/CummingsSNUG2002SJ_FIFO2.pdf}}\BibitemShut
  {NoStop}%
\bibitem [{Note1()}]{Note1}%
  \BibitemOpen
  \bibinfo {note} {\label {note96bits}An optional extended version uses 12
  instead of 8 bytes per sample. This allows to have two independent buses
  driven by a single FPGA-SoC board with a modified buffer card.}\BibitemShut
  {Stop}%
\bibitem [{Note2()}]{Note2}%
  \BibitemOpen
  \bibinfo {note} {The strobe signal is generated by the FPGA. For $\Gamma
  _{sample}$ = 1\protect \tmspace +\thinmuskip {.1667em}MHz it is a 500\protect
  \tmspace +\thinmuskip {.1667em}ns long pulse starting 240\protect \tmspace
  +\thinmuskip {.1667em}ns after the bus has been updated. The bus clock
  frequency must be at least twice the bus output rate to generate the strobe
  signal.}\BibitemShut {Stop}%
\bibitem [{Note3()}]{Note3}%
  \BibitemOpen
  \bibinfo {note} {Cascading two PLL's is not advised, but in our case, we need
  both for dynamic phase shifting. In addition, this allows to use an external
  clock input pin in a different clocking region which would be otherwise
  inaccessible.}\BibitemShut {Stop}%
\bibitem [{not()}]{note_cdc}%
  \BibitemOpen
  \href@noop {} {}\bibinfo {note} {For simplicity, the two cycles delay
  introduced by the clock-domain crossing (CDC) is not shown in Fig.
  \ref{fig:auto-sync}b.}\BibitemShut {Stop}%
\bibitem [{Note4()}]{Note4}%
  \BibitemOpen
  \bibinfo {note} {The actual algorithm to find the phase jump is similar to
  the Bisection method of finding the root of a function.}\BibitemShut {Stop}%
\bibitem [{Note5()}]{Note5}%
  \BibitemOpen
  \bibinfo {note} {Petalinux 2017.4 from Xilinx which is built on Linux kernel
  version 4.9 and is compiled on Ubuntu LTS 18.04.}\BibitemShut {Stop}%
\bibitem [{NI_()}]{NI_Labview_CVI}%
  \BibitemOpen
  \href
  {https://www.ni.com/en-us/shop/software/programming-environments-for-electronic-test-and-instrumentation-category.html#}
  {}\bibinfo {note} {National Instruments Labview and LabWindows/CVI,
  Programming Environments for Electronic Test and Instrumentation.
  \url{https://www.ni.com/en-us/shop/software/programming-environments-for-electronic-test-and-instrumentation-category.html\#}}\BibitemShut
  {NoStop}%
\bibitem [{\citenamefont {Starkey}\ \emph {et~al.}(2013)\citenamefont
  {Starkey}, \citenamefont {Billington}, \citenamefont {Johnstone},
  \citenamefont {Jasperse}, \citenamefont {Helmerson}, \citenamefont {Turner},\
  and\ \citenamefont {Anderson}}]{Starkey2013}%
  \BibitemOpen
  \bibfield  {author} {\bibinfo {author} {\bibfnamefont {P.}~\bibnamefont
  {Starkey}}, \bibinfo {author} {\bibfnamefont {C.}~\bibnamefont {Billington}},
  \bibinfo {author} {\bibfnamefont {S.}~\bibnamefont {Johnstone}}, \bibinfo
  {author} {\bibfnamefont {M.}~\bibnamefont {Jasperse}}, \bibinfo {author}
  {\bibfnamefont {K.}~\bibnamefont {Helmerson}}, \bibinfo {author}
  {\bibfnamefont {L.}~\bibnamefont {Turner}}, \ and\ \bibinfo {author}
  {\bibfnamefont {R.}~\bibnamefont {Anderson}},\ }\href {\doibase
  10.1063/1.4817213} {\bibfield  {journal} {\bibinfo  {journal} {Rev. Sci.
  Instrum.}\ }\textbf {\bibinfo {volume} {84}},\ \bibinfo {pages} {085111}
  (\bibinfo {year} {2013})},\ \bibinfo {note} {see also
  \url{https://labscriptsuite.org/}}\BibitemShut {NoStop}%
\bibitem [{DIO()}]{DIO64}%
  \BibitemOpen
  \href
  {https://www.viewpointusa.com/product/pxi/dio-64-event-detection-control}
  {}\bibinfo {note} {DIO64 PCI I/O board from Viewpoint Systems, Inc. Requires
  Windows XP/7/8 and PCI slot and is no longer available.
  \url{https://www.viewpointusa.com/product/pxi/dio-64-event-detection-control}}\BibitemShut
  {NoStop}%
\bibitem [{Note6()}]{Note6}%
  \BibitemOpen
  \bibinfo {note} {As permanent storage medium the board uses a micro-SD
  (Secure Digital) card which primarily contains the Linux boot loader and boot
  image but can contain additional files and folders and can be used as a hard
  drive. The Linux image is unpacked by the bootloader in a RAM drive, but if
  needed it can also be expanded into a partition of the SD card. Additionally,
  a USB flash drive can be attached to the board for external
  storage.}\BibitemShut {Stop}%
\bibitem [{\citenamefont {Trenkwalder}, \citenamefont {Zaccanti},\ and\
  \citenamefont {Poli}(2021)}]{Zenodo}%
  \BibitemOpen
  \bibfield  {author} {\bibinfo {author} {\bibfnamefont {A.}~\bibnamefont
  {Trenkwalder}}, \bibinfo {author} {\bibfnamefont {M.}~\bibnamefont
  {Zaccanti}}, \ and\ \bibinfo {author} {\bibfnamefont {N.}~\bibnamefont
  {Poli}},\ }\href {https://doi.org/10.5281/zenodo.4893285} {\emph {\bibinfo
  {title} {Data and analysis for ``A flexible \del{FPGA-SoC based }control
  system for atomic, molecular and optical physics experiments''}}}\ (\bibinfo
  {publisher} {Zenodo},\ \bibinfo {year} {2021})\ \bibinfo {note}
  {\url{https://doi.org/10.5281/zenodo.4893285}}\BibitemShut {NoStop}%
\bibitem [{Note7()}]{Note7}%
  \BibitemOpen
  \bibinfo {note} {The interrupts are generated in the PL part and are thus
  directly accessible during the transmission rate measurement without
  involving the CPU.}\BibitemShut {Stop}%
\bibitem [{Note8()}]{Note8}%
  \BibitemOpen
  \bibinfo {note} {On the TX DMA side we observe a delay of about 30 cycles
  between the arrival of the last data out of the FIFO and the TX
  interrupt.}\BibitemShut {Stop}%
\bibitem [{Zyn()}]{Zynq-SDK-performance}%
  \BibitemOpen
  \href
  {https://www.xilinx.com/support/documentation/sw_manuals/xilinx2018_1/ug1145-sdk-system-performance.pdf}
  {}\bibinfo {note} {Xilinx SDK user guide, system performance analysis,
  UG1145, v2018.2,
  \url{https://www.xilinx.com/support/documentation/sw_manuals/xilinx2018_1/ug1145-sdk-system-performance.pdf}}\BibitemShut
  {NoStop}%
\bibitem [{SoC()}]{SoC-performance}%
  \BibitemOpen
  \href
  {https://www.xilinx.com/support/documentation/application_notes/xapp1219-system-performance-modeling.pdf}
  {}\bibinfo {note} {Xilinx System Performance Analysis of an All Programmable
  SoC, XAPP1219 (v1.1) November 5, 2015,
  \url{https://www.xilinx.com/support/documentation/application_notes/xapp1219-system-performance-modeling.pdf}}\BibitemShut
  {NoStop}%
\bibitem [{Note9()}]{Note9}%
  \BibitemOpen
  \bibinfo {note} {The measured $\Gamma _{DMA}$ corresponds to a maximum
  $\Gamma _{sample}$ of 42\protect \tmspace +\thinmuskip {.1667em}MHz
  (28\protect \tmspace +\thinmuskip {.1667em}MHz) for the 8 (12) bytes per
  sample versions. The given rates apply independently for data output and
  input on the bus and for simultaneous output and input (if the bus
  supports).}\BibitemShut {Stop}%
\bibitem [{\citenamefont {{IEEE 802.3ab}}(1999)}]{GigE}%
  \BibitemOpen
  \bibfield  {author} {\bibinfo {author} {\bibnamefont {{IEEE 802.3ab}}},\
  }\href@noop {} {} (\bibinfo {year} {1999}),\ \bibinfo {note} {{G}igabit
  Ethernet, 1000BASE-T with TCP/IP over Ethernet (II) protocol efficiency of
  95\% for 1460 bytes payload per frame of 1538 bytes.}\BibitemShut {Stop}%
\bibitem [{\citenamefont {Endres}\ \emph {et~al.}(2016)\citenamefont {Endres},
  \citenamefont {Bernien}, \citenamefont {Keesling}, \citenamefont {Levine},
  \citenamefont {Anschuetz}, \citenamefont {Krajenbrink}, \citenamefont
  {Senko}, \citenamefont {Vuletic}, \citenamefont {Greiner},\ and\
  \citenamefont {Lukin}}]{Lukin2016}%
  \BibitemOpen
  \bibfield  {author} {\bibinfo {author} {\bibfnamefont {M.}~\bibnamefont
  {Endres}}, \bibinfo {author} {\bibfnamefont {H.}~\bibnamefont {Bernien}},
  \bibinfo {author} {\bibfnamefont {A.}~\bibnamefont {Keesling}}, \bibinfo
  {author} {\bibfnamefont {H.}~\bibnamefont {Levine}}, \bibinfo {author}
  {\bibfnamefont {E.~R.}\ \bibnamefont {Anschuetz}}, \bibinfo {author}
  {\bibfnamefont {A.}~\bibnamefont {Krajenbrink}}, \bibinfo {author}
  {\bibfnamefont {C.}~\bibnamefont {Senko}}, \bibinfo {author} {\bibfnamefont
  {V.}~\bibnamefont {Vuletic}}, \bibinfo {author} {\bibfnamefont
  {M.}~\bibnamefont {Greiner}}, \ and\ \bibinfo {author} {\bibfnamefont
  {M.~D.}\ \bibnamefont {Lukin}},\ }\href {\doibase 10.1126/science.aah3752}
  {\bibfield  {journal} {\bibinfo  {journal} {Science}\ }\textbf {\bibinfo
  {volume} {354}},\ \bibinfo {pages} {1024} (\bibinfo {year} {2016})},\ \Eprint
  {http://arxiv.org/abs/https://science.sciencemag.org/content/354/6315/1024.full.pdf}
  {https://science.sciencemag.org/content/354/6315/1024.full.pdf} \BibitemShut
  {NoStop}%
\bibitem [{\citenamefont {Sahin}\ \emph {et~al.}(2017)\citenamefont {Sahin},
  \citenamefont {Geppert}, \citenamefont {Müllers},\ and\ \citenamefont
  {Ott}}]{Ott2017}%
  \BibitemOpen
  \bibfield  {author} {\bibinfo {author} {\bibfnamefont {C.}~\bibnamefont
  {Sahin}}, \bibinfo {author} {\bibfnamefont {P.}~\bibnamefont {Geppert}},
  \bibinfo {author} {\bibfnamefont {A.}~\bibnamefont {Müllers}}, \ and\
  \bibinfo {author} {\bibfnamefont {H.}~\bibnamefont {Ott}},\ }\href {\doibase
  10.1088/1367-2630/aa9461} {\bibfield  {journal} {\bibinfo  {journal} {New J.
  Phys.}\ }\textbf {\bibinfo {volume} {19}},\ \bibinfo {pages} {123005}
  (\bibinfo {year} {2017})}\BibitemShut {NoStop}%
\bibitem [{Note10()}]{Note10}%
  \BibitemOpen
  \bibinfo {note} {The change in the rate between using a single or two threads
  on both boards is only about 10\%.}\BibitemShut {Stop}%
\bibitem [{Note11()}]{Note11}%
  \BibitemOpen
  \bibinfo {note} {For cable lengths $< 3$\protect \tmspace +\thinmuskip
  {.1667em}m the actual setup cannot detect the round-trip time since the
  reflected pulse is too close to the generated one. However, this situation is
  automatically detected and with the proposed scheme and further technical
  improvements shorter cables should be detectable.}\BibitemShut {Stop}%
\bibitem [{RG5()}]{RG58}%
  \BibitemOpen
  \href {https://www.tasker.it/db_files/products/276044f7e2.pdf} {}\bibinfo
  {note} {{T}asker {RG58 CU} coaxial cable specification gives velocity factor
  0.66, corresponding to a propagation delay of 5.05(4)\,ns/m. See
  \url{https://www.tasker.it/db_files/products/276044f7e2.pdf}}\BibitemShut
  {NoStop}%
\bibitem [{Note12()}]{Note12}%
  \BibitemOpen
  \bibinfo {note} {For the measurement on the secondary board the pulse is not
  reflected to avoid interference of the incoming with the reflected pulse.
  However, we have not observed a difference in the measurement
  result.}\BibitemShut {Stop}%
\bibitem [{USB()}]{USBTMC}%
  \BibitemOpen
  \href
  {https://www.usb.org/document-library/test-measurement-class-specification}
  {}\bibinfo {note} {{USB} test and measurement class (USBTMC).
  \url{https://www.usb.org/document-library/test-measurement-class-specification}}\BibitemShut
  {NoStop}%
\bibitem [{GPI()}]{GPIB}%
  \BibitemOpen
  \href {https://standards.ieee.org/standard/488_2-1992.html} {}\bibinfo {note}
  {{G}eneral purpose interface bus (GPIB), IEEE 488.2.
  \url{https://standards.ieee.org/standard/488_2-1992.html}}\BibitemShut
  {NoStop}%
\bibitem [{\citenamefont {Liu}\ \emph {et~al.}(2018)\citenamefont {Liu},
  \citenamefont {Lü}, \citenamefont {Chen}, \citenamefont {Li}, \citenamefont
  {Qu}, \citenamefont {Wang}, \citenamefont {Li}, \citenamefont {Ren},
  \citenamefont {Dong}, \citenamefont {Zhao}, \citenamefont {Xia},
  \citenamefont {Zhao}, \citenamefont {Ji}, \citenamefont {Ye}, \citenamefont
  {Sun}, \citenamefont {Yao}, \citenamefont {Song}, \citenamefont {Liang},
  \citenamefont {Hu}, \citenamefont {Yu}, \citenamefont {Hou}, \citenamefont
  {Shi}, \citenamefont {Zang}, \citenamefont {Xiang}, \citenamefont {Peng},\
  and\ \citenamefont {Wang}}]{ClockSpace2018}%
  \BibitemOpen
  \bibfield  {author} {\bibinfo {author} {\bibfnamefont {L.}~\bibnamefont
  {Liu}}, \bibinfo {author} {\bibfnamefont {D.-S.}\ \bibnamefont {Lü}},
  \bibinfo {author} {\bibfnamefont {W.-B.}\ \bibnamefont {Chen}}, \bibinfo
  {author} {\bibfnamefont {T.}~\bibnamefont {Li}}, \bibinfo {author}
  {\bibfnamefont {Q.-Z.}\ \bibnamefont {Qu}}, \bibinfo {author} {\bibfnamefont
  {B.}~\bibnamefont {Wang}}, \bibinfo {author} {\bibfnamefont {L.}~\bibnamefont
  {Li}}, \bibinfo {author} {\bibfnamefont {W.}~\bibnamefont {Ren}}, \bibinfo
  {author} {\bibfnamefont {Z.-R.}\ \bibnamefont {Dong}}, \bibinfo {author}
  {\bibfnamefont {J.-B.}\ \bibnamefont {Zhao}}, \bibinfo {author}
  {\bibfnamefont {W.-B.}\ \bibnamefont {Xia}}, \bibinfo {author} {\bibfnamefont
  {X.}~\bibnamefont {Zhao}}, \bibinfo {author} {\bibfnamefont {J.-W.}\
  \bibnamefont {Ji}}, \bibinfo {author} {\bibfnamefont {M.-F.}\ \bibnamefont
  {Ye}}, \bibinfo {author} {\bibfnamefont {Y.-G.}\ \bibnamefont {Sun}},
  \bibinfo {author} {\bibfnamefont {Y.-Y.}\ \bibnamefont {Yao}}, \bibinfo
  {author} {\bibfnamefont {D.}~\bibnamefont {Song}}, \bibinfo {author}
  {\bibfnamefont {Z.-G.}\ \bibnamefont {Liang}}, \bibinfo {author}
  {\bibfnamefont {S.-J.}\ \bibnamefont {Hu}}, \bibinfo {author} {\bibfnamefont
  {D.-H.}\ \bibnamefont {Yu}}, \bibinfo {author} {\bibfnamefont
  {X.}~\bibnamefont {Hou}}, \bibinfo {author} {\bibfnamefont {W.}~\bibnamefont
  {Shi}}, \bibinfo {author} {\bibfnamefont {H.-G.}\ \bibnamefont {Zang}},
  \bibinfo {author} {\bibfnamefont {J.-F.}\ \bibnamefont {Xiang}}, \bibinfo
  {author} {\bibfnamefont {X.-K.}\ \bibnamefont {Peng}}, \ and\ \bibinfo
  {author} {\bibfnamefont {Y.-Z.}\ \bibnamefont {Wang}},\ }\href {\doibase
  10.1038/s41467-018-05219-z} {\bibfield  {journal} {\bibinfo  {journal} {Nat.
  Commun.}\ }\textbf {\bibinfo {volume} {9}},\ \bibinfo {pages} {2760}
  (\bibinfo {year} {2018})}\BibitemShut {NoStop}%
\bibitem [{\citenamefont {Aveline}\ \emph {et~al.}(2020)\citenamefont
  {Aveline}, \citenamefont {Williams}, \citenamefont {Elliott}, \citenamefont
  {Dutenhoffer}, \citenamefont {Kellogg}, \citenamefont {Kohel}, \citenamefont
  {Lay}, \citenamefont {Oudrhiri}, \citenamefont {Shotwell}, \citenamefont
  {Yu},\ and\ \citenamefont {Thompson}}]{BECinspace2020}%
  \BibitemOpen
  \bibfield  {author} {\bibinfo {author} {\bibfnamefont {D.~C.}\ \bibnamefont
  {Aveline}}, \bibinfo {author} {\bibfnamefont {J.~R.}\ \bibnamefont
  {Williams}}, \bibinfo {author} {\bibfnamefont {E.~R.}\ \bibnamefont
  {Elliott}}, \bibinfo {author} {\bibfnamefont {C.}~\bibnamefont
  {Dutenhoffer}}, \bibinfo {author} {\bibfnamefont {J.~R.}\ \bibnamefont
  {Kellogg}}, \bibinfo {author} {\bibfnamefont {J.~M.}\ \bibnamefont {Kohel}},
  \bibinfo {author} {\bibfnamefont {N.~E.}\ \bibnamefont {Lay}}, \bibinfo
  {author} {\bibfnamefont {K.}~\bibnamefont {Oudrhiri}}, \bibinfo {author}
  {\bibfnamefont {R.~F.}\ \bibnamefont {Shotwell}}, \bibinfo {author}
  {\bibfnamefont {N.}~\bibnamefont {Yu}}, \ and\ \bibinfo {author}
  {\bibfnamefont {R.~J.}\ \bibnamefont {Thompson}},\ }\href {\doibase
  10.1038/s41586-020-2346-1} {\bibfield  {journal} {\bibinfo  {journal}
  {Nature}\ }\textbf {\bibinfo {volume} {582}},\ \bibinfo {pages} {193}
  (\bibinfo {year} {2020})}\BibitemShut {NoStop}%
\bibitem [{\citenamefont {Lachmann}\ \emph {et~al.}(2021)\citenamefont
  {Lachmann}, \citenamefont {Ahlers}, \citenamefont {Becker}, \citenamefont
  {Dinkelaker}, \citenamefont {Grosse}, \citenamefont {Hellmig}, \citenamefont
  {Müntinga}, \citenamefont {Schkolnik}, \citenamefont {Seidel}, \citenamefont
  {Wendrich}, \citenamefont {Wenzlawski}, \citenamefont {Carrick},
  \citenamefont {Gaaloul}, \citenamefont {Lüdtke}, \citenamefont {Braxmaier},
  \citenamefont {Ertmer}, \citenamefont {Krutzik}, \citenamefont {Lämmerzahl},
  \citenamefont {Peters}, \citenamefont {Schleich}, \citenamefont {Sengstock},
  \citenamefont {Wicht}, \citenamefont {Windpassinger},\ and\ \citenamefont
  {Rasel}}]{InterferometerSpace2021}%
  \BibitemOpen
  \bibfield  {author} {\bibinfo {author} {\bibfnamefont {M.~D.}\ \bibnamefont
  {Lachmann}}, \bibinfo {author} {\bibfnamefont {H.}~\bibnamefont {Ahlers}},
  \bibinfo {author} {\bibfnamefont {D.}~\bibnamefont {Becker}}, \bibinfo
  {author} {\bibfnamefont {A.~N.}\ \bibnamefont {Dinkelaker}}, \bibinfo
  {author} {\bibfnamefont {J.}~\bibnamefont {Grosse}}, \bibinfo {author}
  {\bibfnamefont {O.}~\bibnamefont {Hellmig}}, \bibinfo {author} {\bibfnamefont
  {H.}~\bibnamefont {Müntinga}}, \bibinfo {author} {\bibfnamefont
  {V.}~\bibnamefont {Schkolnik}}, \bibinfo {author} {\bibfnamefont {S.~T.}\
  \bibnamefont {Seidel}}, \bibinfo {author} {\bibfnamefont {T.}~\bibnamefont
  {Wendrich}}, \bibinfo {author} {\bibfnamefont {A.}~\bibnamefont
  {Wenzlawski}}, \bibinfo {author} {\bibfnamefont {B.}~\bibnamefont {Carrick}},
  \bibinfo {author} {\bibfnamefont {N.}~\bibnamefont {Gaaloul}}, \bibinfo
  {author} {\bibfnamefont {D.}~\bibnamefont {Lüdtke}}, \bibinfo {author}
  {\bibfnamefont {C.}~\bibnamefont {Braxmaier}}, \bibinfo {author}
  {\bibfnamefont {W.}~\bibnamefont {Ertmer}}, \bibinfo {author} {\bibfnamefont
  {M.}~\bibnamefont {Krutzik}}, \bibinfo {author} {\bibfnamefont
  {C.}~\bibnamefont {Lämmerzahl}}, \bibinfo {author} {\bibfnamefont
  {A.}~\bibnamefont {Peters}}, \bibinfo {author} {\bibfnamefont {W.~P.}\
  \bibnamefont {Schleich}}, \bibinfo {author} {\bibfnamefont {K.}~\bibnamefont
  {Sengstock}}, \bibinfo {author} {\bibfnamefont {A.}~\bibnamefont {Wicht}},
  \bibinfo {author} {\bibfnamefont {P.}~\bibnamefont {Windpassinger}}, \ and\
  \bibinfo {author} {\bibfnamefont {E.~M.}\ \bibnamefont {Rasel}},\ }\href
  {\doibase 10.1038/s41467-021-21628-z} {\bibfield  {journal} {\bibinfo
  {journal} {Nat. Commun.}\ }\textbf {\bibinfo {volume} {12}},\ \bibinfo
  {pages} {1317} (\bibinfo {year} {2021})}\BibitemShut {NoStop}%
\bibitem [{\citenamefont {Hensen}\ \emph {et~al.}(2015)\citenamefont {Hensen},
  \citenamefont {Bernien}, \citenamefont {Dréau}, \citenamefont {Reiserer},
  \citenamefont {Kalb}, \citenamefont {Blok}, \citenamefont {Ruitenberg},
  \citenamefont {Vermeulen}, \citenamefont {Schouten}, \citenamefont
  {Abellán}, \citenamefont {Amaya}, \citenamefont {Bruneri}, \citenamefont
  {Mitchell}, \citenamefont {Markham}, \citenamefont {Twitchen}, \citenamefont
  {Elkouss}, \citenamefont {Wehner}, \citenamefont {Taminiau},\ and\
  \citenamefont {Hanson}}]{Hensen2015}%
  \BibitemOpen
  \bibfield  {author} {\bibinfo {author} {\bibfnamefont {B.}~\bibnamefont
  {Hensen}}, \bibinfo {author} {\bibfnamefont {H.}~\bibnamefont {Bernien}},
  \bibinfo {author} {\bibfnamefont {A.~E.}\ \bibnamefont {Dréau}}, \bibinfo
  {author} {\bibfnamefont {A.}~\bibnamefont {Reiserer}}, \bibinfo {author}
  {\bibfnamefont {N.}~\bibnamefont {Kalb}}, \bibinfo {author} {\bibfnamefont
  {M.~S.}\ \bibnamefont {Blok}}, \bibinfo {author} {\bibfnamefont
  {J.}~\bibnamefont {Ruitenberg}}, \bibinfo {author} {\bibfnamefont {R.~F.~L.}\
  \bibnamefont {Vermeulen}}, \bibinfo {author} {\bibfnamefont {R.~N.}\
  \bibnamefont {Schouten}}, \bibinfo {author} {\bibfnamefont {C.}~\bibnamefont
  {Abellán}}, \bibinfo {author} {\bibfnamefont {W.}~\bibnamefont {Amaya}},
  \bibinfo {author} {\bibfnamefont {V.}~\bibnamefont {Bruneri}}, \bibinfo
  {author} {\bibfnamefont {M.~W.}\ \bibnamefont {Mitchell}}, \bibinfo {author}
  {\bibfnamefont {M.}~\bibnamefont {Markham}}, \bibinfo {author} {\bibfnamefont
  {D.~J.}\ \bibnamefont {Twitchen}}, \bibinfo {author} {\bibfnamefont
  {D.}~\bibnamefont {Elkouss}}, \bibinfo {author} {\bibfnamefont
  {S.}~\bibnamefont {Wehner}}, \bibinfo {author} {\bibfnamefont {T.~H.}\
  \bibnamefont {Taminiau}}, \ and\ \bibinfo {author} {\bibfnamefont
  {R.}~\bibnamefont {Hanson}},\ }\href {\doibase 10.1038/nature15759}
  {\bibfield  {journal} {\bibinfo  {journal} {Nature}\ }\textbf {\bibinfo
  {volume} {526}},\ \bibinfo {pages} {682} (\bibinfo {year}
  {2015})}\BibitemShut {NoStop}%
\bibitem [{\citenamefont {Graham}\ \emph {et~al.}(2013)\citenamefont {Graham},
  \citenamefont {Hogan}, \citenamefont {Kasevich},\ and\ \citenamefont
  {Rajendran}}]{Kasevich2013}%
  \BibitemOpen
  \bibfield  {author} {\bibinfo {author} {\bibfnamefont {P.~W.}\ \bibnamefont
  {Graham}}, \bibinfo {author} {\bibfnamefont {J.~M.}\ \bibnamefont {Hogan}},
  \bibinfo {author} {\bibfnamefont {M.~A.}\ \bibnamefont {Kasevich}}, \ and\
  \bibinfo {author} {\bibfnamefont {S.}~\bibnamefont {Rajendran}},\ }\href
  {\doibase 10.1103/PhysRevLett.110.171102} {\bibfield  {journal} {\bibinfo
  {journal} {Phys. Rev. Lett.}\ }\textbf {\bibinfo {volume} {110}},\ \bibinfo
  {pages} {171102} (\bibinfo {year} {2013})}\BibitemShut {NoStop}%
\bibitem [{\citenamefont {Canuel}\ \emph {et~al.}(2018)\citenamefont {Canuel},
  \citenamefont {Bertoldi}, \citenamefont {Amand}, \citenamefont {Pozzo~di
  Borgo}, \citenamefont {Chantrait}, \citenamefont {Danquigny}, \citenamefont
  {Dovale~Álvarez}, \citenamefont {Fang}, \citenamefont {Freise},
  \citenamefont {Geiger}, \citenamefont {Gillot}, \citenamefont {Henry},
  \citenamefont {Hinderer}, \citenamefont {Holleville}, \citenamefont {Junca},
  \citenamefont {Lefèvre}, \citenamefont {Merzougui}, \citenamefont {Mielec},
  \citenamefont {Monfret}, \citenamefont {Pelisson}, \citenamefont
  {Prevedelli}, \citenamefont {Reynaud}, \citenamefont {Riou}, \citenamefont
  {Rogister}, \citenamefont {Rosat}, \citenamefont {Cormier}, \citenamefont
  {Chaibi}, \citenamefont {Gaffet},\ and\ \citenamefont {Bouyer}}]{Bouyer2018}%
  \BibitemOpen
  \bibfield  {author} {\bibinfo {author} {\bibfnamefont {B.}~\bibnamefont
  {Canuel}}, \bibinfo {author} {\bibfnamefont {A.}~\bibnamefont {Bertoldi}},
  \bibinfo {author} {\bibfnamefont {L.}~\bibnamefont {Amand}}, \bibinfo
  {author} {\bibfnamefont {E.}~\bibnamefont {Pozzo~di Borgo}}, \bibinfo
  {author} {\bibfnamefont {T.}~\bibnamefont {Chantrait}}, \bibinfo {author}
  {\bibfnamefont {C.}~\bibnamefont {Danquigny}}, \bibinfo {author}
  {\bibfnamefont {M.}~\bibnamefont {Dovale~Álvarez}}, \bibinfo {author}
  {\bibfnamefont {B.}~\bibnamefont {Fang}}, \bibinfo {author} {\bibfnamefont
  {A.}~\bibnamefont {Freise}}, \bibinfo {author} {\bibfnamefont
  {R.}~\bibnamefont {Geiger}}, \bibinfo {author} {\bibfnamefont
  {J.}~\bibnamefont {Gillot}}, \bibinfo {author} {\bibfnamefont
  {S.}~\bibnamefont {Henry}}, \bibinfo {author} {\bibfnamefont
  {J.}~\bibnamefont {Hinderer}}, \bibinfo {author} {\bibfnamefont
  {D.}~\bibnamefont {Holleville}}, \bibinfo {author} {\bibfnamefont
  {J.}~\bibnamefont {Junca}}, \bibinfo {author} {\bibfnamefont
  {G.}~\bibnamefont {Lefèvre}}, \bibinfo {author} {\bibfnamefont
  {M.}~\bibnamefont {Merzougui}}, \bibinfo {author} {\bibfnamefont
  {N.}~\bibnamefont {Mielec}}, \bibinfo {author} {\bibfnamefont
  {T.}~\bibnamefont {Monfret}}, \bibinfo {author} {\bibfnamefont
  {S.}~\bibnamefont {Pelisson}}, \bibinfo {author} {\bibfnamefont
  {M.}~\bibnamefont {Prevedelli}}, \bibinfo {author} {\bibfnamefont
  {S.}~\bibnamefont {Reynaud}}, \bibinfo {author} {\bibfnamefont
  {I.}~\bibnamefont {Riou}}, \bibinfo {author} {\bibfnamefont {Y.}~\bibnamefont
  {Rogister}}, \bibinfo {author} {\bibfnamefont {S.}~\bibnamefont {Rosat}},
  \bibinfo {author} {\bibfnamefont {A.}~\bibnamefont {Cormier}, \bibfnamefont
  {E.~Landragin}}, \bibinfo {author} {\bibfnamefont {W.}~\bibnamefont
  {Chaibi}}, \bibinfo {author} {\bibfnamefont {S.}~\bibnamefont {Gaffet}}, \
  and\ \bibinfo {author} {\bibfnamefont {P.}~\bibnamefont {Bouyer}},\ }\href
  {\doibase 10.1038/s41598-018-32165-z} {\bibfield  {journal} {\bibinfo
  {journal} {Sci. Rep.}\ }\textbf {\bibinfo {volume} {8}},\ \bibinfo {pages}
  {14064} (\bibinfo {year} {2018})}\BibitemShut {NoStop}%
\bibitem [{\citenamefont {Takamoto}\ \emph {et~al.}(2020)\citenamefont
  {Takamoto}, \citenamefont {Ushijima}, \citenamefont {Ohmae}, \citenamefont
  {Yahagi}, \citenamefont {Kokado}, \citenamefont {Shinkai},\ and\
  \citenamefont {Katori}}]{Katori2020}%
  \BibitemOpen
  \bibfield  {author} {\bibinfo {author} {\bibfnamefont {M.}~\bibnamefont
  {Takamoto}}, \bibinfo {author} {\bibfnamefont {I.}~\bibnamefont {Ushijima}},
  \bibinfo {author} {\bibfnamefont {N.}~\bibnamefont {Ohmae}}, \bibinfo
  {author} {\bibfnamefont {T.}~\bibnamefont {Yahagi}}, \bibinfo {author}
  {\bibfnamefont {K.}~\bibnamefont {Kokado}}, \bibinfo {author} {\bibfnamefont
  {H.}~\bibnamefont {Shinkai}}, \ and\ \bibinfo {author} {\bibfnamefont
  {H.}~\bibnamefont {Katori}},\ }\href {\doibase 10.1038/s41566-020-0619-8}
  {\bibfield  {journal} {\bibinfo  {journal} {Nat. Photonics}\ }\textbf
  {\bibinfo {volume} {14}},\ \bibinfo {pages} {411} (\bibinfo {year}
  {2020})}\BibitemShut {NoStop}%
\bibitem [{Note13()}]{Note13}%
  \BibitemOpen
  \bibinfo {note} {If $\phi _{-} \approx \phi _{+}$ the measurement is not
  reliable due to its sensitivity to noise.}\BibitemShut {Stop}%
\bibitem [{Note14()}]{Note14}%
  \BibitemOpen
  \bibinfo {note} {This choice was motivated to have similar $\protect \frac
  {d\tau _p}{d L}$ for the measurements of the primary and secondary board. The
  average in $\protect \frac {d\tau _p}{d L}$ for the leading and trailing edge
  of the pulse is the same for both boards, but the primary board shows a
  larger discrepancy between the values obtained for the two edges. The
  difference is caused by the dispersion of the pulse. $\phi _0$ corrects the
  phase shift introduced by half of the pulse width $w_p/2$ (see Fig. \ref
  {fig:auto-sync-timing}) but does not correct for the changing width along the
  path.}\BibitemShut {Stop}%
\bibitem [{Note15()}]{Note15}%
  \BibitemOpen
  \bibinfo {note} {The value of $\varphi _p^{crit}$ (see Tab. \ref
  {tab:constants}) has been determined experimentally, but there might be a
  dependence with our choice of parameters. Its exact origin has not been
  investigated.}\BibitemShut {Stop}%
\bibitem [{Note16()}]{Note16}%
  \BibitemOpen
  \bibinfo {note} {We do not use the feedback option which cancels such phase
  shifts.}\BibitemShut {Stop}%
\bibitem [{Note17()}]{Note17}%
  \BibitemOpen
  \bibinfo {note} {When fitting $\sigma $, the correlation to the slope $k$
  causes that for some traces the fit has problems to converge and attains big
  errors.}\BibitemShut {Stop}%
\bibitem [{Note18()}]{Note18}%
  \BibitemOpen
  \bibinfo {note} {In cycling mode the board repeats the experimental sequence
  for a programmed number of times or infinitely until a stop command is
  sent.}\BibitemShut {Stop}%
\end{thebibliography}%

\end{document}